# 1. Uncertainty-aware integration of deuterium metabolic imaging with genome-scale metabolic modeling reveals diet-associated feasible-space remodeling in mouse skeletal muscle

**Tin Yau Pang [1,2,3,†,*], Samia Maouhoub [4,5,†], Maruf Khan [1,2,3], Andreas Korzowski [6], Michael Roden [1,2,7], Ulrich Flögel [4,5,‡], and Daniel Dörr [1,2,3,‡,*]**

1 Department for Endocrinology and Diabetology, Medical Faculty and University Hospital Düsseldorf, Heinrich Heine University Düsseldorf
2 German Diabetes Center (DDZ), Leibniz Institute for Diabetes Research, Germany
3 Center for Digital Medicine, Heinrich Heine University Düsseldorf, Germany
4 Institute for Molecular Cardiology, Medical Faculty and University Hospital Düsseldorf, Heinrich-Heine-University Düsseldorf, Germany
5 Cardiovascular Research Institute Düsseldorf (CARID), Medical Faculty and University Hospital Düsseldorf, Heinrich-Heine University Düsseldorf, Germany
6 , Germany
7 German Center for Diabetes Research, Partner Düsseldorf, München-Neuherberg, Germany
† Equal contributions
‡ Equal contributions
* Correspondence: pang@hhu.de for Tin Yau Pang; daniel.doerr@hhu.de for Daniel Dörr

**Abstract:** Deuterium metabolic imaging (DMI) enables non-invasive measurement of glucose-derived labeling in vivo, but translating noisy, partially identified DMI time courses into network-level metabolic interpretation remains computationally challenging. We present an uncertainty-aware DMI-to-GEM workflow that converts dynamic $^{2}H$ metabolic imaging data into soft constraints for genome-scale metabolic modeling and uses feasible-space analysis to propagate these constraints through a skeletal-muscle metabolic network. Following intraperitoneal [6,6′-$^{2}H_{2}$]-glucose administration, dynamic $^{2}H$ spectra were acquired from Gastrocnemius, Soleus, and Tibialis anterior in control and high-fat diet (HFD) mice. A hierarchical four-state kinetic model was fitted to estimate glucose-derived metabolic parameters while accounting for shared plasma input, tissue-specific parameters, voxel-level fluxes, kinetic isotope effects, and metabolite-dependent uncertainty. Because glucose labeling was more robust than lactate and glutamate/glutamine labeling, only two DMI-derived anchors were imposed as direct GEM inputs: glucose-processing rate was mapped to hexokinase flux, and glucose-label-derived oxidative entry was mapped to mitochondrial pyruvate dehydrogenase flux. These anchors were implemented as two-sided soft target intervals with weighted mismatch penalties, defining one DMI-compatible feasible region per mouse–tissue replicate. Flux variability analysis and flux sampling were then used to quantify reaction- and subsystem-level feasible-space remodeling. The DMI-derived constraints were stoichiometrically compatible with the reconstructed muscle model and propagated beyond the constrained reactions to alter feasible glycolytic, lactate-associated, pyruvate oxidation, TCA-cycle, oxidative phosphorylation, and redox-associated activity. HFD-associated remodeling was tissue dependent, with Tibialis anterior and Soleus showing reduced glucose-linked feasible activity, whereas Gastrocnemius showed a mixed response with preserved or increased TCA/redox-associated feasible activity. This work establishes DMI-informed genome-scale modeling as a computational framework for integrating non-invasive tracer imaging with network-level flux reasoning while avoiding overinterpretation of weakly identified metabolite signals.

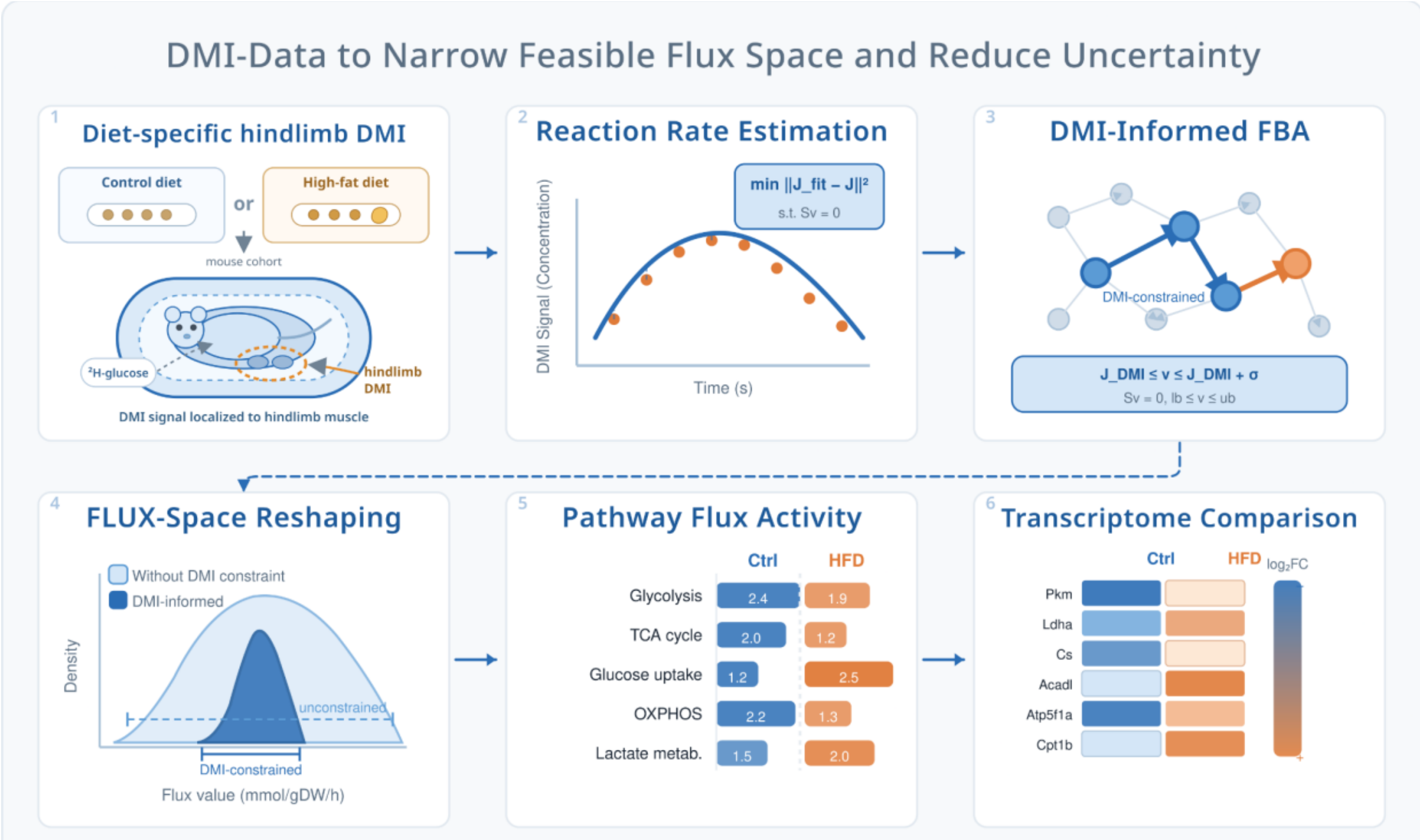

**Figure 1. Overview of the deuterium metabolic imaging (DMI)-informed metabolic modeling framework.** DMI data acquired from mouse hindlimb were used to estimate physiological parameters and infer metabolic flux constraints through model fitting and optimization. The inferred fluxes were incorporated into a genome-scale metabolic model (GEM) using hierarchical FBA to generate DMI-informed metabolic flux states. FVA and flux sampling were subsequently performed to characterize changes in the feasible metabolic solution space induced by DMI-derived constraints. Pathway and subsystem activities were compared between control and high-fat diet (HFD) conditions, and the resulting FBA-predicted changes in metabolic flux states were evaluated against transcriptomic pathway expression profiles to assess biological consistency.

## 1. Introduction

Flux Balance Analysis (FBA) predicts steady-state metabolic fluxes through genome-scale metabolic model (GEM) by optimizing an objective function (e.g., biomass production or ATP generation) subject to constraints such as stoichiometry or thermodynamics. While powerful for exploring metabolic capabilities, standard FBA relies heavily on assumptions about cellular uptakes, objectives, and reaction bounds. To improve biological relevance, numerous methods integrate transcriptomic data to further constrain reaction fluxes, for example, by up- or down-regulating the upper/lower bounds according to gene expression levels. Approaches such as TRFBA, LBFBA, and COMPASS have demonstrated that transcriptomics-informed FBA can better predict metabolic phenotypes under different conditions [1–3]. Nevertheless, transcript levels are only indirect proxies for actual enzyme activities and reaction rates. Post-transcriptional regulation, allosteric effects, and substrate availability often decouple gene expression from actual fluxes. Non-invasive measurements of in vivo metabolic fluxes are therefore highly valuable to provide physiologically grounded anchors or to refine these models [4].

Deuterium Metabolic Imaging (DMI) has recently emerged as a non-radioactive magnetic resonance approach for mapping tissue metabolism in vivo. By administering deuterium-labeled substrates such as [6,6′-$^{2}H_2$]-glucose (Glc) and acquiring spatially resolved dynamic $^{2}H$ MR spectra, DMI enables spatiotemporal tracking of substrate uptake and conversion into downstream metabolites such as lactate (Lac), glutamate/glutamine (Glx), and water in vivo. Compared to already established approaches using (also non-radioactive) $^{13}C$-labelled tracers, DMI offers several advantages: (i) Versus conventional $^{13}C$ MRS detection of $^{13}C$-labeled substrates, it exhibits superior sensitivity due to favorable relaxation times that enable very fast data acquisition; furthermore no power-depositing decoupling is required for $^{2}H$; (ii) Although hyperpolarized $^{13}C$ MR techniques are even more sensitive, data acquisition is restricted to 1 or 2 minutes, while DMI allows detection of $^{2}H$-labeled substrates over a period of approximately 2 hours, which enables also for the acquisition of metabolic data from organs which are perfused slowly (e.g, the bone marrow); (iii) Finally, bolus injection of $^{13}C$-labeled substrates and subsequent collection of blood samples for analysis by mass spectrometry is limited to analyze metabolites in the blood and provides only indirect information about organ metabolism. In contrast, kinetic modeling of the time courses obtained by DMI yields quantitative estimates of metabolic rates (e.g., glycolytic flux) directly in living tissue, including skeletal muscle [5,6]. In the context of genome-scale modeling, DMI-derived flux estimates could provide functional constraints for GEMs, but their use is non-trivial because downstream metabolite signals are noisy, partially identified, and substrate-specific rather than direct measurements of total pathway flux. Unlike transcriptomics alone, DMI provides real flux estimates that can tighten feasible solution spaces, validate model predictions, or guide hierarchical optimization strategies. This integration bridges non-invasive human imaging with systems-level GEM, potentially overcoming limitations of purely transcriptomics/proteomics-driven or in silico approaches.

It remains unclear how DMI-derived flux estimates—especially when noisy and only partially identified—can be propagated through GEMs without overinterpreting them as direct pathway measurements. Addressing this challenge requires a framework that explicitly represents measurement uncertainty and maps it onto the feasible solution space of the metabolic network, rather than treating DMI-derived fluxes as hard constraints. Here, we adapted DMI-based kinetic modeling [7,8] to investigate skeletal muscle metabolism in a high-fat diet (HFD) mouse model (**Fig. 1**). DMI data acquired from mouse hindlimb were used to estimate physiological parameters and infer metabolic flux constraints through kinetic model fitting and optimization. These uncertainty-weighted DMI-derived fluxes were then incorporated into a skeletal muscle GEM, and flux variability analysis (FVA) and flux sampling were performed to characterize diet-induced changes in the feasible metabolic solution space. The methodological contribution of this study is an uncertainty-aware DMI-to-GEM integration strategy, in which imaging-derived kinetic estimates are not treated as exact flux measurements, but as soft, reliability-weighted anchors that define replicate-specific feasible regions for downstream FVA and flux sampling.

## 2. Materials and Methods

### *2.1. Animal and DMI acquisition*

Male C57BL/6J mice (9–12 weeks) were assigned to control (n = 4) or high-fat diet (HFD; n = 7) groups and imaged longitudinally at baseline, week 3, week 6, and week 9. Dynamic, spatially resolved $^2$H MR spectra of the hindlimb were acquired at 9.4T with a 8-mm $^1$H/$^2$H solenoid and a 2D CSI-SSFP sequence (TR = 350 ms, flip angle = 60°, max 90 min) following intraperitoneal injection of [6,6′-$^2H_2$]-glucose at 2 mg/g body weight. Full acquisition details are provided in **Supp. Text S1**.

### *2.2. Spectral Processing and ROI Annotation*

Three hindlimb muscle regions—Gastrocnemius, Soleus, and Tibialis anterior—were delineated using eemanually selected seed voxels expanded to a 3×3 neighborhood and restricted by a tissue mask excluding bone and subcutaneous adipose tissue (see **Supp. Text S2** and **Fig. 2**). Metabolite-specific signals were extracted by integrating preprocessed $^2$H MR spectra within fixed chemical shift windows: glucose (3.3–3.9 ppm), lactate (1.1–1.5 ppm), and glutamate/glutamine Glx (2.0–2.7 ppm). Spectral preprocessing details including ppm correction, baseline definition, and smoothing are described in **Supp. Text S2**.

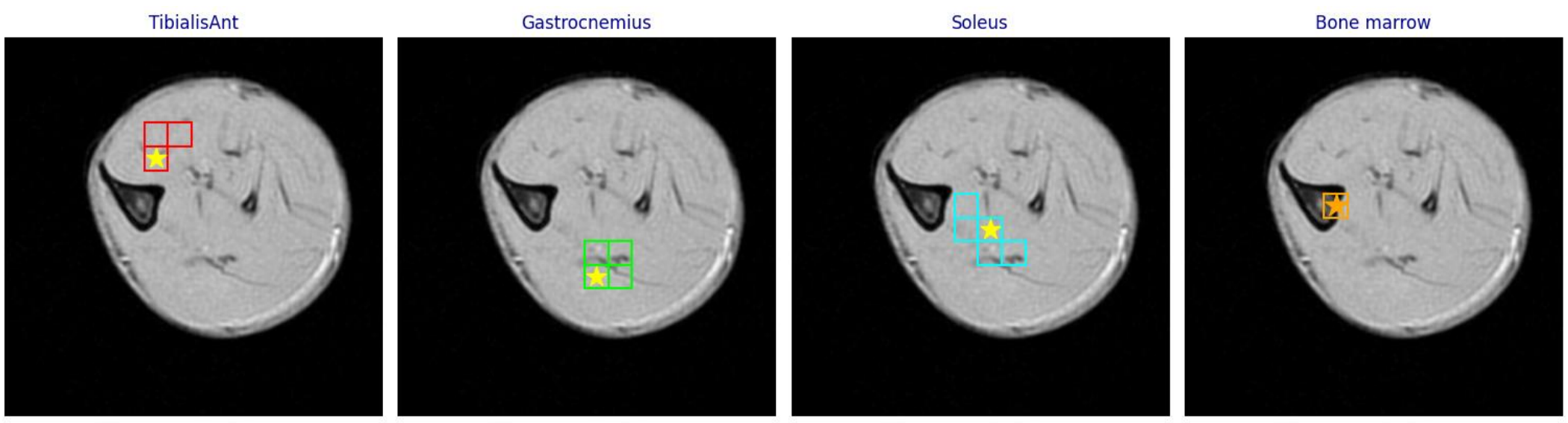

**Figure 2. Representative hindlimb ROI annotation.** Cross-sectional $^1$H MRI from a week-9 HFD mouse showing manually seeded regions for Tibialis anterior, Gastrocnemius, Soleus, and Bone marrow. Seed voxels were expanded to local 3 × 3 neighborhoods and restricted by tissue masking before extraction of dynamic $^2$H glucose, lactate, and Glx signals. Bone marrow was used as a reference region and excluded from muscle flux modeling.

### *2.3. Deriving Metabolic Fluxes from DMI Data.*

We estimated glucose-derived metabolic flux descriptors by fitting a four-state ordinary differential equation (ODE) model to the observed dynamic $^2$H time courses of glucose, lactate, and glutamate/glutamine (Glx) (**Fig. 3**). The model was adapted from previous DMI kinetic modeling studies [7,8] and modified for skeletal-muscle interpretation. It tracks labeled intracellular glucose, a labeled lactate/pyruvate pool, a labeled $\alpha$-ketoglutarate (aKG) pool, and a downstream Glx pool. The model is driven by a subject-specific plasma $^2$H-glucose input function, represented as a delayed biexponential curve with finite rise time (**Fig. 4**). Because plasma $^2$H-glucose was not measured directly, one shared plasma input function was fitted jointly across all available muscle tissues for each mouse-week.

The fitted parameters were organized hierarchically. Plasma input parameters were shared across tissues within each mouse-week; tissue-level parameters included glucose-transport capacity ratio $\eta$, $\alpha$-ketoglutarate–Glx exchange $V_x$, and blood-volume fraction $f_v$; and voxel-level parameters described glucose processing $MR_{glc}$, lactate/pyruvate label clearance $V_{lac,loss}$, glucose-label-derived oxidative entry $V_{ox}$, and the glucose-derived fraction of effective TCA/Glx turnover $\varphi_{TCA}$. The Michaelis transport constant was fixed at 5.8 mM [9]. Kinetic isotope effect and deuterium label-loss corrections were applied using previously reported values [10]. Model predictions were averaged over the DMI acquisition window before comparison with observations, corresponding to 280 s per time point. Full equations, fixed constants, parameter bounds, loss function, and optimization settings are provided in **Supp. Text S3** and **Supp. Table S1**.

We used the fitted glucose metabolic rate, $MR_{glc}$, and glucose-label-derived oxidative entry, $V_{ox}$ in the downstream analysis. Total effective TCA/Glx turnover was derived as $V_{TCA,total} = V_{ox}/\varphi_{TCA}$, and the apparent dilution component was calculated as $V_{dil,TCA} = V_{TCA,total} - V_{ox}$. Because DMI measures glucose-derived label appearance rather than total pathway flux, $MR_{glc}$ and $V_{ox}$ were treated as the most reliable DMI-visible flux anchors for genome-scale modeling. Lactate- and Glx-dependent quantities, including $V_{lac,loss}$, $V_{TCA,total}$, $V_{dil,TCA}$, and $\varphi_{TCA}$, were retained for diagnostic description rather than hard constraints on individual GEM reactions. This conservative interpretation was used to avoid overinterpreting weaker downstream Lac* and Glx* signals.

**DMI Glucose-to-Lactate/TCA ODE System**

$$\mathrm{Glc}_p^*(t) = a_1\left(1 - e^{-(t-\delta)/\tau_{\mathrm{rise}}}\right)\left[w e^{-k_{\mathrm{fast}}(t-\delta)} + (1-w)e^{-k_{\mathrm{slow}}(t-\delta)}\right]$$

$$T_{\max} = \eta\,\mathrm{MR}_{\mathrm{glc}}, \quad V_{\mathrm{TCA,\,total}} = V_{\mathrm{ox}}/\phi_{\mathrm{TCA}}$$

$$\frac{d\mathrm{Glc}^*}{dt} = \frac{T_{\max}\mathrm{Glc}_p^*}{K_T + \mathrm{Glc}_p + \mathrm{Glc}_p^*} - \frac{T_{\max}\mathrm{Glc}^*}{K_T + \mathrm{Glc}_c + \mathrm{Glc}^*} - \frac{\mathrm{MR}_{\mathrm{glc}}}{\mathrm{KIE}_{\mathrm{gly}}}\frac{\mathrm{Glc}^*}{\mathrm{Glc}_c + \mathrm{Glc}^*}$$

$$\frac{d\mathrm{Lac}^*}{dt} = (1 - f_{\mathrm{gly}})\frac{\mathrm{MR}_{\mathrm{glc}}}{\mathrm{KIE}_{\mathrm{gly}}}\frac{\mathrm{Glc}^*}{\mathrm{Glc}_c + \mathrm{Glc}^*} - V_{\mathrm{lac,\,loss}}\frac{\mathrm{Lac}^*}{\mathrm{Lac} + \mathrm{Lac}^*} - \frac{V_{\mathrm{ox}}}{\mathrm{KIE}_{\mathrm{tca}}}\frac{\mathrm{Lac}^*}{\mathrm{Lac} + \mathrm{Lac}^*}$$

$$\frac{d\mathrm{aKG}^*}{dt} = (1 - f_{\mathrm{tca}})\frac{V_{\mathrm{ox}}}{\mathrm{KIE}_{\mathrm{tca}}}\frac{\mathrm{Lac}^*}{\mathrm{Lac} + \mathrm{Lac}^*} - (V_{\mathrm{TCA,\,total}} + V_x)\frac{\mathrm{aKG}^*}{\mathrm{aKG} + \mathrm{aKG}^*} + V_x\frac{\mathrm{Glx}^*}{\mathrm{Glx} + \mathrm{Glx}^*}$$

$$\frac{d\mathrm{Glx}^*}{dt} = V_x\left(\frac{\mathrm{aKG}^*}{\mathrm{aKG} + \mathrm{aKG}^*} - \frac{\mathrm{Glx}^*}{\mathrm{Glx} + \mathrm{Glx}^*}\right)$$

**Figure 3. DMI kinetic model used to estimate glucose-derived fluxes.** The four-state ODE model tracks labeled glucose (Glc), lactate (Lac), $\alpha$-ketoglutarate (aKG), and Glx after intraperitoneal [6,6′-$^2H_2$]-glucose administration. Glucose transport is modeled with Michaelis–Menten kinetics, while fitted parameters describe glucose utilization, lactate partitioning, apparent glucose-label-derived TCA-associated turnover, lactate washout, and aKG–Glx exchange. Kinetic isotope effects and label-loss corrections were applied before mapping fitted fluxes to the GEM.

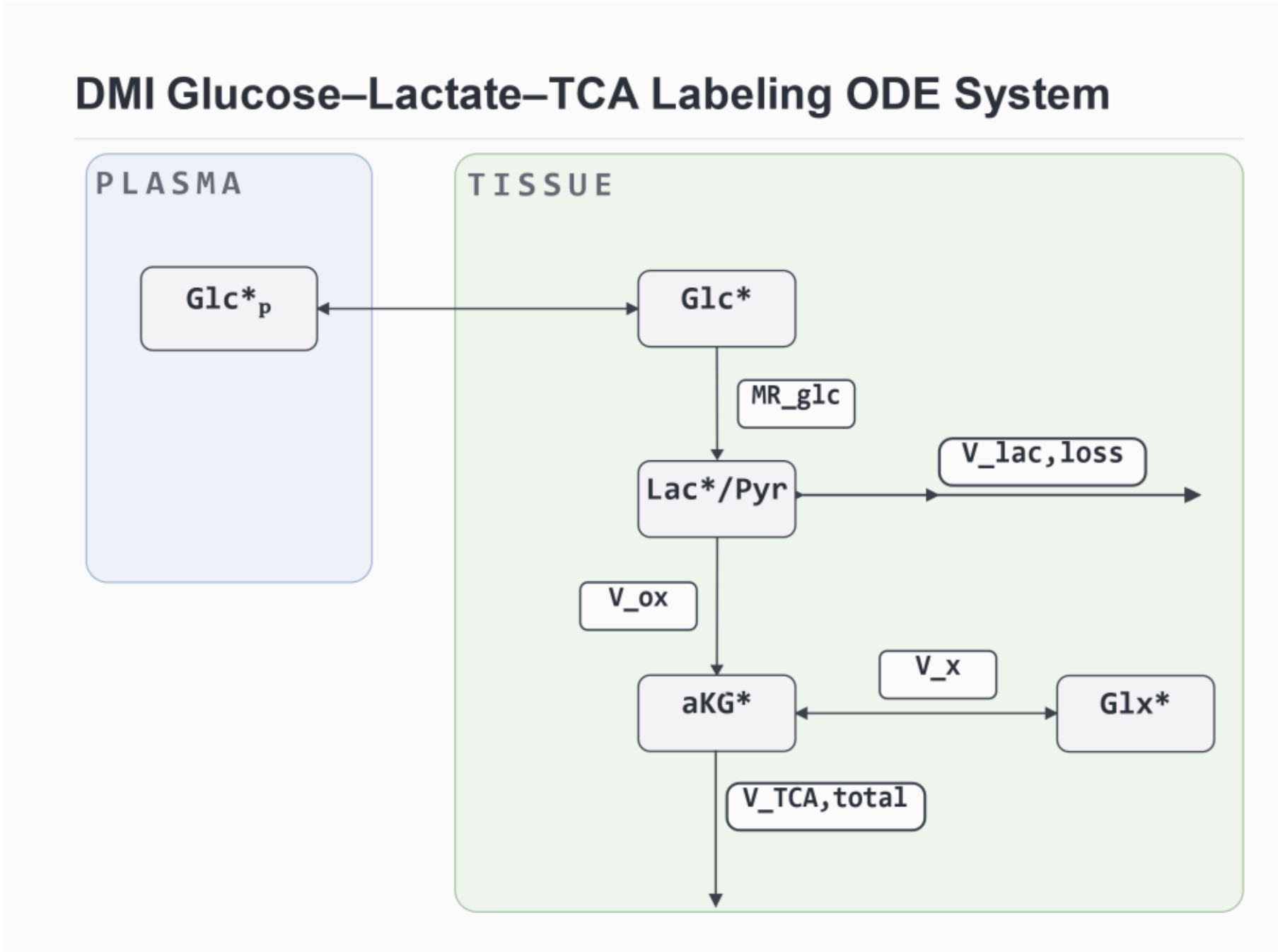

**Figure 4. Compartmental representation of labeled metabolite flow in hindlimb muscle.** The model follows labeled glucose, lactate/pyruvate, aKG, and Glx, driven by a fitted delayed-rise biexponential plasma $^2$H-glucose input function. Glucose, lactate, and Glx are directly observed by DMI, whereas the intermediate pool links lactate/pyruvate oxidation to downstream Glx labeling. $V_{ox}$ is interpreted as DMI-visible glucose-label-derived oxidative entry, while $V_{TCA,total} = V_{ox}/\varphi_{TCA}$ represents a derived total effective TCA/Glx turnover diagnostic rather than a measured TCA-cycle flux.

*2.4. GEM Construction and DMI-Constrained FBA*

We reconstructed a mouse skeletal muscle-specific genome-scale metabolic model from the iMM1865 mouse GEM [11] using CORDA [12]. Normalized transcriptomic data from GSE17576 [13] were used to assign gene-confidence scores, which were propagated to reactions through gene-protein-reaction rules. Key muscle-relevant reactions required for downstream simulations, including mitochondrial ATP synthase and biomass maintenance, were protected during reconstruction. The final muscle-specific model contained 7,653 reactions, 4,913 metabolites, and 1,745 genes. Full reconstruction details are provided in **Supp. Text S5**.

We performed DMI-constrained simulations in COBRApy [14] using one model instance per week-9 mouse–tissue replicate. The model was simulated under a defined muscle medium allowing uptake of glucose, oxygen, hexadecanoate, beta-hydroxybutyrate, and water, with a fixed ATP hydrolysis demand imposed through DM_atp_c_. The current analysis used a conservative DMI-to-GEM mapping (**Table 1**; **Supp. Table S2**): $MR_{glc}$ was mapped to hexokinase flux (HEX1) as a glucose-processing constraint, and $V_{ox}$ was mapped to mitochondrial pyruvate dehydrogenase flux (PDHm) as a glucose-label-derived oxidative pyruvate-entry constraint. Lactate- and TCA-dilution-related quantities were retained for diagnosis and biological interpretation but were not imposed as GEM constraints.

Because the DMI kinetic model and GEM use different units, DMI-derived rates were converted to GEM flux units using a single global scale factor derived from the week-9 control $MR_{glc}$ reference set, such that the average control $MR_{glc}$ corresponded to a reference HEX1 flux of 0.10 mmol $gDW^{-1}$ $h^{-1}$. This scaling was used to preserve relative differences among mice, tissues, and diet groups rather than to claim absolute biochemical flux calibration; therefore, absolute GEM flux magnitudes should be interpreted cautiously, and scale factor sensitivity should be evaluated in future analyses. Constraints were imposed as soft target intervals rather than exact equalities, allowing uncertainty-aware deviations when required for model feasibility. For each replicate, the model minimized a weighted mismatch penalty for deviations outside the DMI-derived target intervals, and the relaxed optimal penalty defined the final DMI-constrained feasible region. Flux variability analysis and flux sampling were then used to characterize reaction-level flux ranges and subsystem activity distributions. Details of the exchange-bound policy, scaling, soft-constraint formulation, penalty relaxation, sampling, and metabolic subsystem (i.e., pathway) scoring are provided in **Supp. Text S6**.

**Table 1. DMI-to-GEM constraint mapping.** DMI-derived flux quantities and their corresponding reactions and constraint types in the muscle-specific genome-scale model. Full biological descriptions and FBA interpretations are provided in **Supp. Table S2**.

| DMI quantity | FBA reaction | Constraint type | Interpretation |
|---|---|---|---|
| $MR_{glc}$ | HEX1 | Two-sided soft match | Glucose-processing / intracellular glucose phosphorylation anchor |
| $V_{ox}$ | PDHm | Two-sided soft match | Glucose-label-derived oxidative pyruvate-entry anchor |
| $V_{lac,loss}$ | Not constrained | Diagnostic only | Lactate/pyruvate label clearance or compatibility aliases; not mapped to LDH_L |
| $V_{TCA,total}$, $V_{dil,TCA}$, $\varphi_{TCA}$ | Not constrained | Diagnostic only | Dilution/TCA-associated C6 quantities; not interpreted as direct total TCA-cycle constraints |

*2.5. Pathway Activity Scoring and Transcriptomic Comparison*

Subsystem-level flux activity was summarized by averaging the absolute flux of all reactions assigned to each metabolic subsystem (i.e., pathway) in the GEM, yielding pathway activity scores for each constrained state. To contextualize model predictions, reaction-associated genes were mapped to the same subsystems and gene-module activity scores were computed from two independent public skeletal-muscle transcriptomic datasets (GSE17576 and GSE305719). These transcriptomic scores were used as contextual support for biological plausibility, not as matched-sample validation; full details of gene mapping and score computation are in **Supp. Text S7**.

*2.6. Precautionary modeling strategies against overinterpretation of DMI-derived constraints*

Because DMI-derived kinetic parameters differ in identifiability and do not necessarily correspond to complete biochemical pathway fluxes, we used a conservative DMI-to-GEM mapping strategy to avoid overinterpretation. Only the most reliable glucose-derived quantities, $MR_{glc}$ and $V_{ox}$, were imposed as direct soft GEM constraints, whereas lactate- and Glx-dependent quantities were retained for diagnostic interpretation. In addition, DMI-derived targets were implemented as soft intervals rather than exact equality constraints, and FVA/flux-sampling results were interpreted as feasible model variation conditional on DMI constraints, nutrient bounds, ATP demand, and GEM assumptions. The main precautionary strategies are summarized in **Table 2**, with implementation details and suggested sensitivity analyses provided in **Supp. Table S3**.

**Table 2. Precautionary modeling strategies used in the DMI-to-GEM workflow.**

| Potential reviewer concern | Precautionary modeling strategies used in this study | Remaining limitation |
|---|---|---|
| Downstream Lac* and Glx* DMI signals are noisier than Glc*. | Only the more robust DMI-derived quantities, $MR_{glc}$ and $V_{ox}$, were used as GEM constraints. Lac/Glx-dependent quantities were retained as diagnostic descriptors. | Lactate- and Glx-associated fitted quantities remain less identifiable and should not be interpreted as direct pathway fluxes. |
| DMI-derived rates are not equivalent to total biochemical pathway fluxes. | $MR_{glc}$ was interpreted as a glucose-processing anchor and $V_{ox}$ as glucose-label-derived oxidative pyruvate entry, not as complete glycolytic or TCA-cycle flux. | Absolute flux magnitudes depend on DMI model assumptions, correction factors, and scaling to GEM units. |
| Hard constraints could make the GEM infeasible or overfit noisy estimates. | DMI inputs were imposed as two-sided soft target intervals with mismatch penalties, allowing deviation when required for stoichiometric feasibility. | Choice of tolerance and penalty weights should be tested in future sensitivity analyses. |
| The MRI-to-GEM unit conversion is approximate. | A single global scale factor was applied across all replicates to preserve relative mouse, tissue, and group differences. | The scale factor is not an independently measured physical conversion and should be stress-tested. |
| FBA sampling distributions may look like biological replicates. | Sampling was interpreted as feasible model variation conditional on DMI constraints, not as independent biological replication. | Biological inference remains limited by the number of mice and the assumptions of the GEM. |
| Lactate exchange in skeletal muscle can be bidirectional. | $V_{lac,loss}$ was not mapped directly to LDH_L or lactate export, reducing risk of incorrect lactate interpretation. | The current ODE model does not explicitly include plasma lactate uptake, lactate enrichment, or lactate oxidation. |
| Transcriptomic comparison is not matched to the DMI cohort. | RNA modules were used only as contextual biological support, not as direct validation. | Matched transcriptomic, metabolomic, biochemical, or isotope-tracing validation is needed for stronger biological conclusions. |

## 3. Results

### *3.1. DMI kinetic fitting captures glucose-labeling dynamics but reveals metabolite-dependent uncertainty*

We first evaluated whether the compartmental DMI kinetic model could reproduce the observed dynamic $^2H$-labeling time courses in hindlimb skeletal muscle before using the fitted quantities as inputs for genome-scale modeling. Across the fitted mouse-week cohorts, model agreement was strongest for labeled glucose, where predicted and observed Glc* concentrations closely followed the identity line and yielded a pooled $R^2$ of 0.86 (**Fig. 5**). In contrast, the downstream metabolite channels showed substantially weaker agreement, with pooled $R^2$ values of 0.17 for Lac* and -0.22 for Glx*. The per-voxel $R^2$ distributions confirmed this pattern: Glc* fits were generally reliable, whereas Lac* fits were more variable and Glx* fits frequently showed poor voxel-level agreement. Thus, the kinetic model was primarily constrained by glucose-labeling dynamics, while Lac* and especially Glx* provided weaker and more noise-sensitive information.

Representative voxel-wise fits illustrate the same behaviour at the tissue level (**Fig. 6**). The model reproduced the main temporal accumulation and clearance pattern of Glc* across Tibialis anterior, Gastrocnemius, and Soleus, whereas Lac* and Glx* showed larger voxel-to-voxel variability and less consistent agreement with the fitted curves. We did not observe an obvious group-specific fitting bias in the pooled observed-versus-predicted plots; rather, the dominant limitation was metabolite-specific signal quality, with downstream Lac* and Glx* estimates being less tightly identified than Glc*. This likely reflects the lower signal-to-noise ratio, baseline sensitivity, and partial spectral overlap of downstream resonances compared with the administered glucose signal.

Uncertainty diagnostics supported cautious interpretation of the fitted fluxes. The mouse-week fits were classified as boundary-sensitive or weakly identified, indicating that the fits were usable for group-level and tissue-level summaries but that individual mouse-level parameter estimates should not be overinterpreted (**Supp. Fig. S1**). The draw-based confidence intervals were also broader for some fitted quantities, especially those depending on downstream Lac*/Glx* dynamics. For this reason, we treated the DMI-derived fluxes as uncertainty-aware model inputs rather than direct measurements of complete pathway fluxes. In the subsequent GEM analysis, only $MR_{glc}$ and $V_{ox}$ were imposed as soft constraints, whereas $V_{lac,loss}$, $V_{TCA,total}$, $V_{dil,TCA}$, and $\varphi_{TCA}$ were retained mainly as diagnostic or interpretive quantities, consistent with the DMI-to-GEM mapping described in **Table 1** and **Supp. Table S2**.

Despite these limitations, the week-9 fitted flux summaries were physiologically coherent at the group level (**Supp. Fig. S2** and **Supp. File S1**). Glucose-processing rates were within a similar range across hindlimb muscles, while downstream oxidative and TCA-associated quantities showed tissue- and diet-dependent shifts rather than uniform global changes. In the current week-9 summary, the most notable descriptive trends were reduced $V_{ox}$ and $V_{TCA,total}$ in HFD Tibialis anterior, a smaller reduction of $V_{TCA,total}$ in HFD Soleus, and a modest increase of oxidative/TCA-associated quantities in HFD Gastrocnemius. These trends should be interpreted as exploratory because the sample size was small and several confidence intervals were broad or bound-sensitive. Nevertheless, the fitted values were sufficiently structured and biologically plausible to support their use as soft, uncertainty-aware constraints for the downstream genome-scale metabolic simulations.

Overall, these results show that the current DMI fitting workflow provides reliable glucose-derived dynamic information and usable, but less certain, downstream metabolic descriptors. Therefore, the following FBA analyses were designed to propagate the strongest DMI-derived anchors through a stoichiometrically constrained muscle metabolic network, while avoiding overinterpretation of weakly identified Lac*/Glx*-dependent quantities.

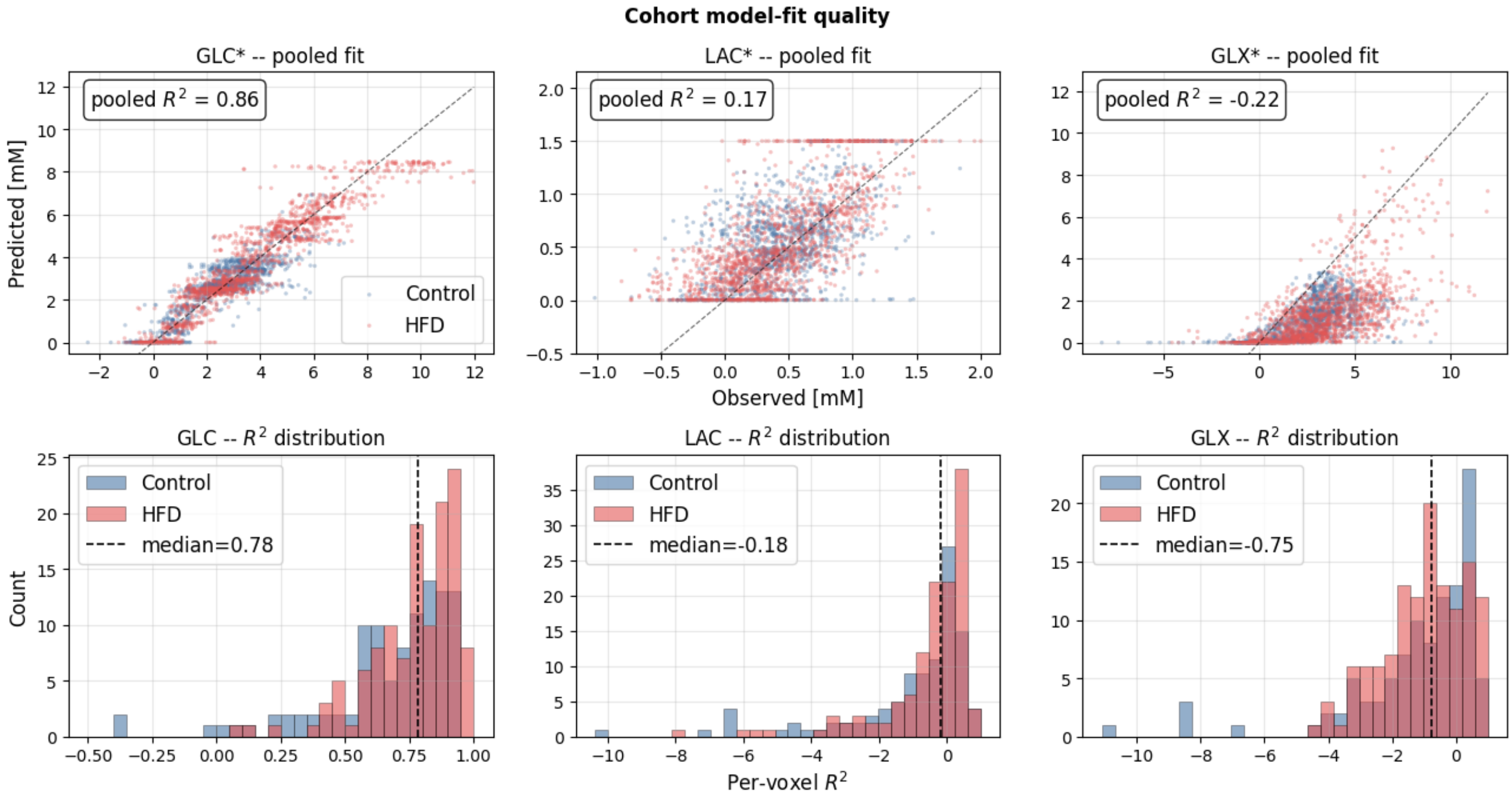

**Figure 5. Cohort model-fit quality across the three measured metabolites.** (Top row) Pooled predicted vs. observed concentrations for glucose (Glc*), lactate (Lac*), and the combined glutamate/glutamine pool (Glx*), with each point representing one timepoint from one voxel of one fit (mM). Markers are coloured by group—Control (blue) vs. HFD (red)—and the dashed black line indicates the identity y = x. The annotated pooled $R^2$ aggregates residuals across all timepoints, voxels, and mice for that metabolite. (Bottom row) Distribution of per-voxel $R^2$ values (one entry per fit × voxel) for each metabolite, stacked by group, with the cohort-wide median shown as a dashed black line. Together, the panels summarise both absolute fit accuracy (top) and the spread of fit quality across individual voxels (bottom) for the cohort of fits retained for downstream analysis.

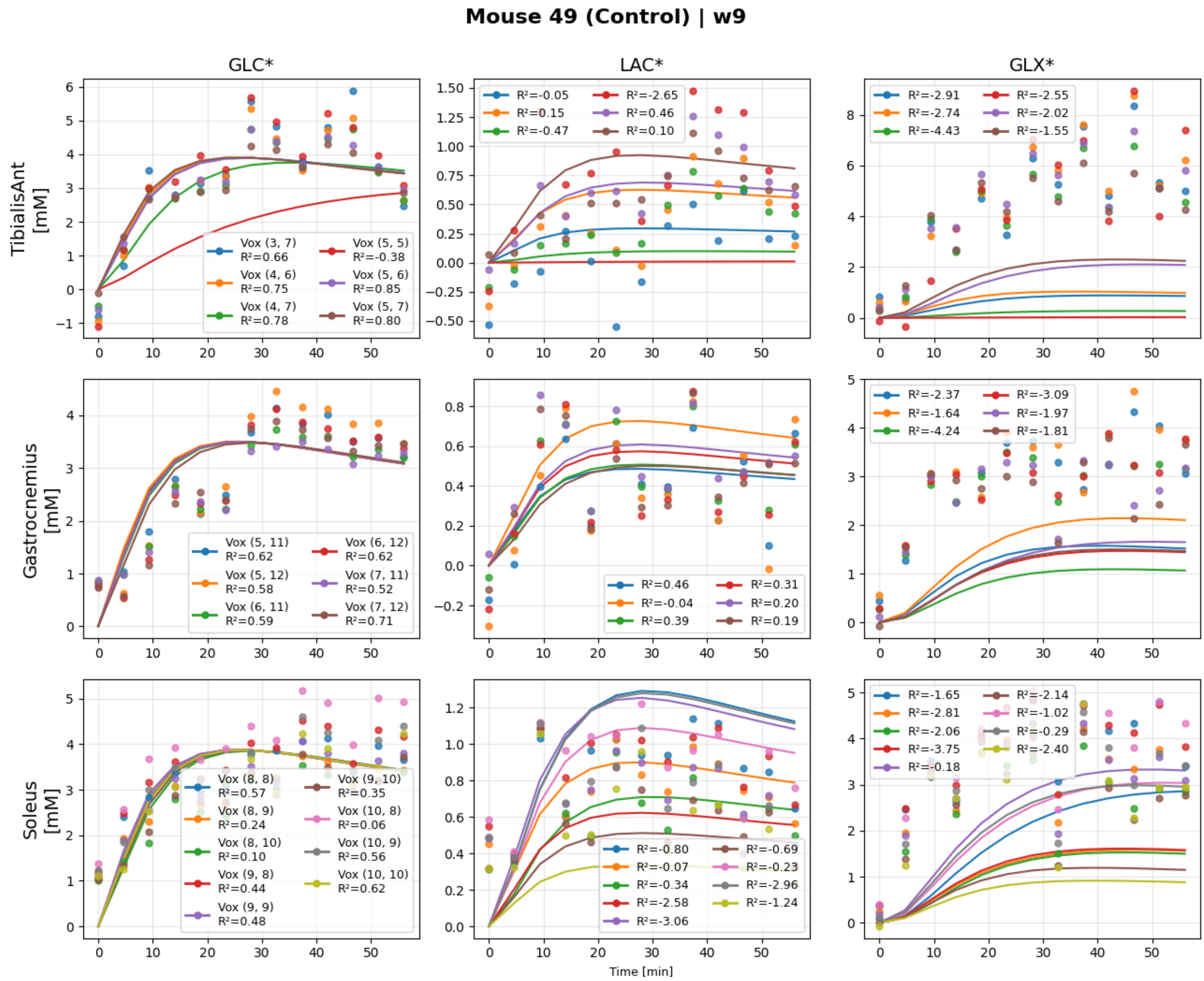

**Figure 6. Representative DMI time-course fits for hindlimb skeletal muscles of an HFD mouse (week 9).** Each row corresponds to one tissue ROI: Tibialis Anterior (top), Gastrocnemius (middle), and Soleus (bottom). Each column shows the temporal accumulation of one deuterium-labeled metabolite following an intraperitoneal [6,6'-$^2H_2$]-glucose bolus: Glc* (left), Lac* (centre), and Glx* (right). Symbols denote concentration derived from measured DMI signal intensities; solid curves are the corresponding fits from an ODE-based compartmental kinetic model in which glucose transport follows Michaelis–Menten kinetics with the transport constant $K_T$ fixed at 5.8 mM [9]. Each colour represents one spatial voxel within the tissue ROI; the coefficient of determination ($R^2$) for each voxel is reported in the legend. A single shared plasma input function is fitted jointly across all three tissues and metabolites simultaneously.

### *3.2. DMI-derived flux constraints reshape the feasible metabolic solution space*

We next tested whether the DMI-derived flux anchors could be embedded into the skeletal-muscle GEM and whether these constraints altered the surrounding feasible flux space. For each week-9 mouse–tissue replicate, the model was constrained using two DMI-derived anchors: $MR_{glc}$ was mapped to hexokinase flux (HEX1) as a glucose-processing constraint, and $V_{ox}$ was mapped to mitochondrial pyruvate dehydrogenase flux (PDHm) as a glucose-label-derived oxidative pyruvate-entry constraint. Lactate- and TCA-dilution-related fitted quantities were retained as diagnostic variables rather than GEM constraints. The imposed-versus-realized comparison showed close agreement between the DMI-derived targets and the corresponding sampled FBA fluxes (**Fig. 7A**), indicating that the DMI-derived soft constraints were compatible with model stoichiometry and could be satisfied without major conflict.

Applying these constraints changed the structure of the feasible metabolic solution space relative to the generic unconstrained reference model. Subsystem-level spread analysis showed that DMI-constrained sampling narrowed the feasible range of oxidative phosphorylation and fatty-acid uptake/beta-oxidation, whereas several connected subsystems showed similar or broader feasible spread, including lactate metabolism, redox shuttles, ketone metabolism, glutamate/aKG metabolism, and TCA-cycle activity (**Fig. 7B**). Thus, the DMI anchors did not simply reduce model uncertainty uniformly. Instead, a small number of DMI-derived constraints redistributed feasible flux variability across connected metabolic pathways. The reference bands of the unconstrained GEM model should therefore be interpreted only as feasible-space references, not as biological confidence intervals or measured control distributions.

DMI-constrained subsystem activity distributions showed tissue-specific HFD-associated remodeling (**Supp. Fig. S3**). Replicate-level summaries showed directionally consistent HFD-associated downward shifts in Tibialis anterior and Soleus, particularly for glucose uptake/phosphorylation, glycolysis, lactate metabolism, and pyruvate oxidation, although exact permutation tests were not significant after summarizing at the mouse-replicate level (**Supp. Table S4**). Gastrocnemius showed a mixed response with reduced lactate/pyruvate-associated activity but preserved or increased TCA-cycle and oxidative-phosphorylation-associated feasible activity. Thus, the dominant HFD effect was not a uniform reduction in mitochondrial ATP production, but a tissue-dependent remodeling of how glucose-derived carbon could be coupled to lactate handling and mitochondrial entry. Because each comparison contained only a small number of mouse-level replicates, exact permutation tests were used as descriptive directional support rather than confirmatory hypothesis tests; the resulting P values were therefore interpreted cautiously.

Reaction-level summaries supported the subsystem-level interpretation (**Supp. Fig. S4**). In addition to the directly anchored HEX1 and PDHm reactions, HFD-associated shifts propagated to glycolytic reactions such as PFK and PYK, the sign-adjusted lactate-associated LDH_L flux, TCA-cycle reactions including CSm and AKGDm, electron-transport reactions such as SUCD1m and CYOR_u10mi, and ATP synthase. These results indicate that DMI-derived glucose and oxidative-labeling anchors influence connected reactions beyond the two directly constrained targets.

Overall, the DMI-constrained FBA analysis suggests that week-9 HFD muscle is characterized by tissue-dependent remodeling of feasible central carbon and mitochondrial metabolism. Tibialis anterior and Soleus showed the most directionally consistent HFD-associated reduction in glucose/phosphorylation, glycolysis, lactate metabolism, and pyruvate oxidation, whereas Gastrocnemius showed a more mixed response with reduced pyruvate/lactate-associated activity but preserved or increased TCA/redox-associated feasible activity. These distributions represent model-implied feasible metabolic states conditional on the DMI-derived constraints, not direct biochemical flux measurements. Nevertheless, they show that even a limited set of DMI-derived anchors can reshape network-level metabolic predictions in a structured, tissue-specific, and biologically interpretable manner.

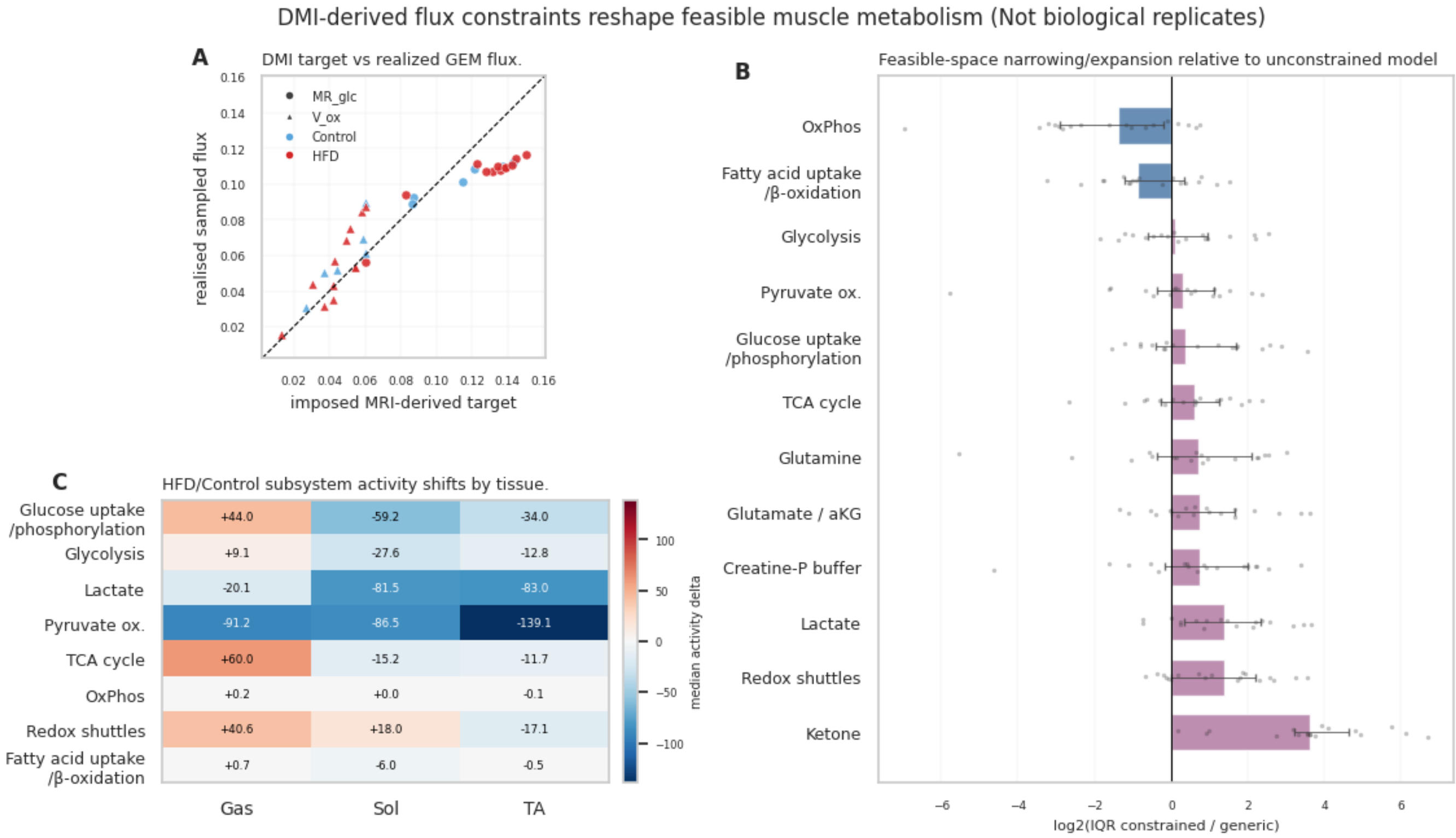

**Figure 7. DMI-derived soft constraints reshape the feasible metabolic solution space of the skeletal-muscle GEM.** (A) Agreement between DMI-derived target fluxes and realized sampled GEM fluxes for the two directly constrained reactions. The glucose-processing rate, $MR_{glc}$, was mapped to hexokinase flux, HEX1, and glucose-label-derived oxidative entry, $V_{ox}$, was mapped to mitochondrial pyruvate dehydrogenase flux, PDHm. Points represent mouse–tissue week-9 replicates, and the dashed line indicates identity. Close agreement indicates that the DMI-derived soft constraints were stoichiometrically compatible with the reconstructed skeletal-muscle model. (B) Constraint-induced changes in feasible subsystem activity spread. Bars show the median $\log_2$ ratio of subsystem interquartile range after DMI constraint integration relative to the generic unconstrained reference model. Negative values indicate narrowing of the feasible activity range, whereas positive values indicate broader feasible variation after applying DMI-derived constraints. Points represent individual mouse–tissue constrained models. (C) Median HFD/Control differences in DMI-constrained subsystem activity across Gastrocnemius, Soleus, and Tibialis anterior. Values summarize sampled feasible-space shifts conditional on the imposed DMI anchors, nutrient bounds, ATP demand, and GEM stoichiometry. These values should be interpreted as model-implied feasible metabolic changes, not as direct biochemical flux measurements or independent biological replicate statistics.

### *3.3. DMI/FBA flux patterns are only weakly mirrored by RNA-based pathway modules*

To evaluate the DMI-constrained FBA predictions, we compared three linked readouts for each metabolic axis: the fitted DMI-derived quantity, the corresponding GEM reaction flux, and an RNA-based module score calculated from genes associated with the same reaction or pathway module. This comparison used two independent skeletal-muscle transcriptomic datasets, GSE17576 [13] for control versus HFD (**Supp. Fig. S5**) and GSE305719 [15] for HFD time-course and tissue context (**Supp. Fig. S6**). Because these datasets were not matched to the DMI cohort by animal, tissue sampling, or experimental protocol, they served as contextual support rather than direct validation.

The strongest DMI–FBA agreement was observed for the directly constrained axes: $MR_{glc}$ was reflected by sampled HEX1 flux, and $V_{ox}$ was reflected by sampled PDHm flux. They were consistent with the imposed DMI-to-GEM mapping. In contrast, the lactate-associated LDH_L flux remained as diagnostic, because $V_{lac,loss}$ was not imposed as a LDH_L constraint. This distinction is important: agreement between $MR_{glc}$/HEX1 and $V_{ox}$/PDHm confirms that the direct anchors were propagated as intended, whereas LDH_L reflects how the constrained network accommodated lactate-associated flux without being directly forced by the DMI fit.

Across both transcriptomic comparisons, RNA module activity showed weaker diet- or tissue-associated separation than the DMI-derived and FBA-sampled flux distributions. Glycolysis-, lactate-, and PDH-associated RNA modules were generally more compact and less variable than the corresponding DMI/FBA readouts. This mismatch is biologically plausible because transcript abundance reflects enzyme-associated regulatory capacity, whereas flux also depends on substrate availability, energetic demand, post-transcriptional regulation, allosteric control, compartmentalization, and network constraints.

Overall, the cross-modal comparison supports two conclusions. First, the direct DMI-to-FBA mappings behaved as expected, with HEX1 tracking $MR_{glc}$ and PDHm tracking $V_{ox}$. Second, RNA module activity did not mirror the DMI/FBA flux-level patterns, suggesting that the integrated workflow captures functional and network-constrained metabolic variation that is not reducible to transcriptional module activity alone. These comparisons should therefore be interpreted as biological context for the model predictions, not as direct validation or falsification of the DMI-constrained FBA results.

**4. Discussion**

In this study, we developed a workflow for integrating DMI-derived metabolic flux estimates with a mouse skeletal-muscle GEM. The central aim was not to treat DMI-derived quantities as direct measurements of complete pathway fluxes, but to use them as uncertainty-aware anchors that restrict and reshape the feasible metabolic solution space. The DMI kinetic model reproduced glucose-labeling dynamics most reliably, whereas lactate and Glx fits were more variable. This metabolite-dependent uncertainty was propagated into the downstream modeling strategy by using only $MR_{glc}$ and $V_{ox}$ as GEM constraints, while lactate- and TCA-dilution-related quantities were interpreted as diagnostic descriptors. The resulting DMI-constrained models remained feasible, matched the imposed glucose-processing and glucose-derived oxidative-entry anchors, and propagated these constraints into broader changes in glycolysis, lactate metabolism, pyruvate oxidation, TCA-associated metabolism, oxidative phosphorylation, and other relevant pathways. More broadly, our work can be viewed as a step toward MRI-compatible fluxomics. Classical isotope-assisted fluxomics and metabolic flux analysis aim to infer reaction rates from tracer dynamics, thereby moving beyond static metabolite abundance toward functional metabolic activity [16,17].

The comparison with transcriptomic modules further illustrates the distinction between expression-based and flux-based readouts. The directly constrained FBA reactions behaved as expected: HEX1 broadly followed $MR_{glc}$, and PDHm followed $V_{ox}$, whereas LDH_L remained a diagnostic feasible-space readout rather than a directly constrained lactate flux. In contrast, RNA module activity was more compact and showed weaker diet- or tissue-associated separation than the DMI-derived and FBA-sampled flux distributions. This partial mismatch is plausible. Transcript abundance reflects enzyme-associated regulatory capacity, whereas flux is also shaped by substrate availability, energetic demand, post-transcriptional regulation, allosteric control, compartmentalization, and network constraints. Thus, agreement between RNA modules and DMI/FBA predictions can support biological plausibility, but weak RNA variation should not be interpreted as evidence that functional flux variation is absent.

Several limitations should be considered. First, downstream lactate and Glx quantification remains the major technical bottleneck. Glc* was the most robust DMI signal, whereas Lac* and Glx* were more sensitive to baseline correction, spectral overlap, low signal-to-noise ratio, and fixed integration windows. This limited the identifiability of lactate- and Glx-dependent kinetic parameters and justified the conservative choice to constrain only $MR_{glc}$ and $V_{ox}$ directly. Second, the present ODE model remains incomplete for skeletal-muscle physiology because it represents lactate/pyruvate label production and apparent clearance, but does not explicitly model plasma lactate uptake, bidirectional lactate exchange, or lactate oxidation as an independent substrate source. This is important because skeletal muscle can both release and consume lactate depending on fiber type, workload, perfusion, and metabolic state [18–20]. More glycolytic regions may export lactate, whereas more oxidative regions may import and oxidize circulating lactate. Third, the plasma glucose and lactate input functions were not measured directly; instead, the plasma $^2$H-glucose input was estimated jointly across tissues for each mouse-week. Direct measurements of plasma glucose enrichment, plasma lactate concentration, and plasma lactate enrichment would improve parameter identifiability and help distinguish local glucose-to-lactate production from circulating lactate uptake.

Additional limitations arise from the small sample size and the modeling assumptions. The FBA sampling distributions are not independent biological replicates; they represent feasible model variation conditional on the imposed DMI constraints, nutrient bounds, ATP demand, and network assumptions. Kinetic isotope effect and deuterium label-loss corrections were taken from prior DMI literature and may not transfer perfectly from brain to skeletal muscle. The muscle GEM was reconstructed using public transcriptomic data and may not capture all condition-specific enzyme capacities in the DMI cohort. Finally, the transcriptomic datasets used for comparison were independent of the DMI cohort and should be interpreted as contextual support rather than matched validation. These issues mean that the current framework is best viewed as hypothesis-generating rather than confirmatory.

Despite these limitations, our study demonstrates that robust DMI-derived flux estimates can be embedded into a stoichiometrically constrained skeletal-muscle GEM and propagated into interpretable, tissue-specific metabolic predictions. The main value of the framework is that it connects non-invasive metabolic imaging to network-level metabolic reasoning while preserving uncertainty and avoiding overinterpretation of weakly identified DMI quantities. Future work should focus on improving lactate and Glx spectral quantification, measuring plasma glucose and lactate input functions, extending the kinetic model to include bidirectional lactate exchange and lactate oxidation, and validating predicted pathway shifts with matched biochemical, metabolomic, or isotope-tracing measurements. With these improvements, DMI-informed GEM analysis could become a useful framework for linking imaging-visible tracer dynamics to systems-level flux interpretation in metabolic disease.

## 5. Conclusions

In summary, we present an uncertainty-aware workflow that uses robust DMI-derived glucose-processing and oxidative-entry estimates as soft constraints in a skeletal-muscle GEM. The resulting feasible spaces reproduced the imposed DMI constraints and suggested tissue-specific HFD-associated remodeling of glucose-linked central carbon metabolism, with stronger reductions in Tibialis anterior and Soleus and a mixed TCA/redox-associated response in Gastrocnemius. These predictions should be interpreted as model-implied feasible states rather than direct biochemical flux measurements. In future work, we should improve lactate and Glx spectral quantification, measure plasma glucose and lactate input functions, and extend the kinetic model to include bidirectional lactate exchange.

**Author Contributions:** DMI-Flux model conceptualization, T.P., M.R. and D.D.; DMI-flux model analysis, T.P. and M.K.; DMI experiment conceptualization, A.K. and U.F.; DMI experimental works: S.M. and U.F.; writing—original draft preparation, T.P. and M.K.; writing—review and editing, T.P., A.K, U.F., M.R. and D.D.

**Funding:** This work was supported by the Deutsche Forschungsgemeinschaft (DFG, German Research Foundation; Grant INST 208/764-1 FUGG to U.F.).

**Acknowledgement:** We like to thank Elton Montrazi and Lucio Frydman from the Weizmann Institute for providing us with the CSI-SSFP sequence used for DMI. Martin Lercher provided computational resources to support this study. Throughout this study, from model simulations and preparation of this manuscript, the authors used Gemini / Grok / Claude / OpenAI GPT-based tools to assist writing the codes, debugging, language editing, restructuring of paragraphs, and generation of suggestions for manuscript organization. AI-assisted output was reviewed, edited, and verified by the authors, who take full responsibility for the content of the manuscript.

**Data Availability Statement:** The data and code related to this work can be found at https://github.com/codialab/DMI_GEM_APBC2026.

**Conflicts of Interest:** The authors declare no conflict of interest.

**Ideally, the main text (not including Abstract, Figure Legends, Author Contributions, Acknowledgments, and References) should be limited to 5,000 words.**

# Supplementary Text

## S1. Animal and DMI acquisition

Experiments were conducted using male C57BL/6J mice aged 9-12 weeks at the study initiation. All MR experiments were performed in a vertical 9.4T Bruker AVANCE$^{NEO}$ NMR spectrometer (Bruker) driven by ParaVision 360v3.6 and operating at a frequency of 400.21 MHz for $^{1}H$ and 61.43 MHz for $^{2}H$. Data were acquired using Bruker Microimaging units (Mini 0.5) and actively shielded 57-mm gradient sets (0.3 T/m maximum gradient strength, 110 µs rise time at 100% gradient switching). The animals were anesthetized with 1.5% isoflurane in air and to prevent drying of the mucous membranes, the anesthetic gas mixture was moistened. The Deuterium metabolic imaging data (DMI) were acquired with an 8-mm $^{1}H/^{2}H$ solenoid and a 2D CSI-SSFP sequence from the mouse hindlimb across four experimental time points: basal, week3, week6 and week 9. Mice were either assigned to the high-fat diet group (HFD; mouse IDs 45, 47, 48, 53, 54, 55, and 56) or control diet (mouse IDs 49, 50, 51, and 52) group. The first acquired CSI-SSFP scan was treated as the baseline frame for each dynamic acquisition.

The experiment was designed to measure the changes of dynamic deuterium signal in hindlimb tissue after intraperitoneal administration of [6,6′-$^{2}H_2$] glucose. In this experimental framework, glucose mainly represents the administered substrate, where lactate reflects glycolytic or anaerobic labeling, and Glx reflects labeling related to TCA cycle. The imaging was performed using dual-coil setup. The $^{1}H$ component was used for planning and anatomical reference imaging, shimming and coil positioning, while the component $^{2}H$ was actively tuned to the deuterium resonance frequency for the detection of the metabolite signals. The acquisition workflow consisted of localizer imaging, map and shim, FLASH reference imaging, a baseline CSI-SSFP acquisition, [6,6′-$^{2}H_2$] glucose administration, CSI-SSFP acquisitions. Each of the CSI-SSFP scans lasted 4 min 40 s, and the dynamic imaging was performed for a max of 90 min. [6,6′-$^{2}H_2$] glucose was administered intraperitoneally at 2 mg/g body weight in a total injection volume of 0.2 mL. The CSI-SSFP acquisition used a repetition time of 350 ms, a flip angle of 60°, a field-of-view of 8×8 $mm^2$, a matrix of 16×16, and a slice thickness of 4 mm. Because metabolite-specific T1 values at 9.4T were unavailable, T1 correction used literature-derived 7T values as an approximation.

## S2. Spectral Processing and ROI Annotation

### S2.1 Hindlimb ROI Annotation

Hindlimb tissue regions were defined using manually selected ROI seed coordinates. Four different anatomical regions were annotated: Bone Marrow, Gastrocnemius (gas), Soleus (sol), and Tibialis anterior (tib). In the downstream muscle analysis, Gastrocnemius, Soleus, and Tibialis anterior were treated as the fitted skeletal muscle tissues (Fig. 2). Bone marrow was retained as a reference region and was not used as a fitted skeletal muscle compartment in the metabolic modelling workflow. One manual seed voxel was originally defined for each anatomical region, mouse, and week. For voxel support and ODE/FBA alignment, the 8 nearest neighboring voxels were used around each seed, corresponding to up to 3 × 3 = 9 candidate voxels, then restricted by tissue mask and signal readiness filtering. The tissue mask is defined to exclude the bones and subcutaneous adipose tissue, whose high lipid content can contaminate the lactate measurement.

### S2.2 Spectral Preprocessing

Raw DMI data were stored as voxelwise spectra files. The ppm axis was corrected using the fixed offset of -5.5 ppm. In the preprocessing workflow, this correction was applied as *ppm=dat(:,1) - ppm_offset.* The same workflow also defined the fixed metabolite windows and scan timing used for the construction of the map. Spectra were smoothed using a Savitzky-Golay filter with polynomial order 3 and window length 11 when sufficient spectra samples were available.

### S2.3 Metabolite Signal Quantification

Metabolite specific signals were quantified from fixed ppm windows. Glucose was extracted from 3.3-3.9 ppm, lactate over 1.1-1.5, whereas the Glx-associated resonance was extracted from 2.0-2.7 ppm. The same ppm correction, fixed metabolite windows, baseline definition, scan timing, and manually defined ROI coordinates were used for all analyses. For each voxel and scan, metabolite signal was quantified by numerical integration of the spectral intensity within the corresponding fixed ppm window after ppm correction and spectral smoothing.

## S3. DMI Kinetic Model and Parameter Estimation

### S3.1 Compartmental ODE Model

Our kinetic model tracks the time evolution of four 2H-labeled metabolite pools in skeletal muscle:

Glc* = intracellular labeled glucose

Lac/Pyr* = labeled lactate/pyruvate pool

K* = labeled TCA-associated intermediate pool

Glx* = labeled glutamate/glutamine pool

The model is adapted from brain DMI literature [1–3] with modifications for skeletal muscle physiology and for dilution-aware interpretation of glucose-derived oxidative labeling. Concentrations are expressed in mM and fluxes in umol/g/min throughout.

The plasma 2H-glucose input is represented by Glc_p*(t).

The labeled-pool fraction function is defined as:

$$\text{pool_fraction}(X^*, X) = X^* / (X + X^*)$$

The fitted glucose and oxidative fluxes are corrected for kinetic isotope effects as:

$$\text{MR_glc_iso} = \text{MR_glc} / \text{kie_gly}$$

$$\text{V_ox_iso} = \text{V_ox} / \text{kie_tca}$$

The glucose-transport capacity is derived from the tissue-level multiplier η:

$$\text{T_max} = \eta * \text{MR_glc}$$

The total effective TCA/Glx turnover is derived as:

$$\text{V_TCA_total} = \text{V_ox} / \text{phi_tca}$$

The apparent dilution component is:

$$V_dil_TCA = V_TCA_total - V_ox$$

The ODE system is shown in **Fig. 3**. The first equation describes labeled glucose transport from plasma into tissue, reversible intracellular glucose transport, and loss of intracellular glucose label through metabolism. The second equation describes production of labeled lactate/pyruvate from glucose label, apparent lactate/pyruvate label clearance or dilution, and oxidative transfer of label into the TCA-associated pool. The third equation describes glucose-label-derived oxidative entry into the TCA-associated intermediate pool (aKG), total effective TCA/Glx turnover, and exchange with Glx. The fourth equation describes reversible exchange between the intermediate pool and Glx. The model is initialized with zero labeled enrichment in all pools.

As in the previous implementation, the glycogen branch of the brain DMI model [1–3] was omitted to reduce model complexity and because glycogen turnover is not directly identifiable from the present DMI acquisition.

## S3.2 Plasma Input Function

The plasma $^2$H-glucose concentration Glc*_p(t) is modeled as a delayed biexponential with a finite sigmoidal rise:

$$Glc^*_p(t) = a_1 \cdot [1 - \exp(-t_eff / \tau_rise)] \cdot [w \cdot \exp(-k_p{,}fast \cdot t_eff) + (1-w) \cdot \exp(-k_p{,}slow \cdot t_eff)]$$

where $t_eff = t - \Delta t_delay$ (set to zero for $t_eff \leq 0$). The biexponential decay captures fast vascular clearance and slower redistribution/metabolism of label. The model has six free parameters: $a_1$, Δt_delay, τ_rise, w, k_p,fast, and k_p,slow. To improve numerical conditioning, k_p,fast is parameterized as k_p,slow + k_p,gap, enforcing k_p,fast > k_p,slow at all times during optimization.

Plasma parameters are shared across all three muscles (tibialis anterior, gastrocnemius, soleus) for a given animal and timepoint — i.e., a single plasma curve is fitted jointly across tissues — reflecting the common arterial input. Tissue-specific blood volume fractions ($f_v \in [0, 0.15]$) account for partial volume contamination of the $^2$H-glucose signal by intravascular label.

DMI signal at each timepoint integrates over the full ~4.7-minute acquisition window. Model predictions are therefore computed as the mean of the simulated time course over the corresponding window interval (9 interpolated points across the scan window on each nominal timepoint), rather than as instantaneous values.

## S3.3 Fixed Constants, Kinetic Isotope Effects, and Label-Loss Corrections

The fixed pool sizes and correction constants used in the simulations can be found **Supp. Table S1**. The KIE values kie_gly = 1.042 and kie_tca = 1.035 follow de Graaf et al. [4]. Deuterium label-loss correction factors f_gly = 0.157 and f_tca = 0.385 represent label loss before the measured downstream pools are reached. Their use follows prior DMI modeling, but they may not transfer perfectly to skeletal muscle and are therefore a source of systematic uncertainty in absolute flux estimates.

## S3.4 Parameter Hierarchy and Bounds

Parameters are organized into three levels:

Global (shared across all tissues per animal/timepoint): six plasma input parameters ($a_1$, $\Delta t_delay$, $\tau_rise$, w, k_p,slow, k_p,gap). These are estimated jointly in Phase 2.

Tissue-level (shared across voxels within a muscle): η, V_x (α-KG ↔ Glx exchange flux), and f_v (blood volume fraction). Three parameters per tissue × three tissues = 9 tissue-level parameters in Phase 2.

Voxel-level: MR_glc, V_lac,loss, V_ox, φ_TCA. Four parameters per voxel.

The Michaelis transport constant K_T is fixed at K_T = 5.8 mM [3].

### S3.5 Loss Function

For each voxel and metabolite channel, glucose, lactate, and Glx, a reliability score is computed from dynamic range, positive time points, and signal-to-noise ratio. The noise level sigma is estimated from the median absolute deviation of first differences, with a minimum floor of 0.05 mM.

Standardized residuals are:

r_i = (y_pred_i - y_obs_i) / sigma

Residuals are scored using the pseudo-Huber loss:

pseudo_huber(r, delta) = delta^2 * [sqrt(1 + (r / delta)^2) - 1]

with:

delta = 1.5

Point weights up-weight positive-signal time points, and glucose receives additional weight during the first 15 min of the time course. Trend losses based on first differences are included with weights:

w_trend_Glc = 0.75

w_trend_Glx = 0.35

w_trend_Lac = 0.20

The Glx trend term is applied only when Glx reliability exceeds 0.35, and the lactate trend term only when lactate reliability exceeds 0.50. Secondary channels, lactate and Glx, are excluded from a voxel's objective if their reliability falls below 0.15.

The metabolite losses are aggregated as reliability-weighted means across voxels and channels. A weak quadratic regularization penalty is added for plasma parameters and for f_v, with f_v regularized toward a prior of 0.02 using scale 0.03 and weight 0.03.

### S3.6 Optimization Strategy

Fitting proceeds in two phases.

Phase 1 fits each mouse-week-tissue combination independently. The Phase 1 shared parameter vector contains 9 parameters: η, V_x, a1, delta_t_delay, tau_rise, w, k_p_slow, k_p_gap, f_v. Each voxel contributes 4 additional parameters: MR_glc, V_lac_loss, V_ox, phi_tca.

Phase 1 optimization uses differential evolution with Latin hypercube initialization, population size 8, maximum 150 iterations, and tolerance 1e-5, followed by L-BFGS-B refinement with maximum 2000 iterations, f_tol = 1e-9, and g_tol = 1e-6. If L-BFGS-B fails to converge, a single perturbed retry is attempted using Gaussian noise with sigma = 1e-3.

Phase 2 jointly fits all available tissues for each mouse-week using one shared plasma input. Each tissue has its own eta, V_x, and f_v, and each voxel has its own MR_glc, V_lac_loss, V_ox, and phi_tca. Phase 1 solutions are used as warm starts, with plasma parameters initialized as the mean across Phase 1 tissue fits. Phase 2 differential evolution uses population size 8 and 50-100 iterations, followed by the same L-BFGS-B refinement.

The Phase 2 parameter vector has the following structure:

6 global plasma parameters + 3 parameters per tissue+ 4 parameters per voxel

or equivalently:

$$N_parameters = 6 + 3 * N_tissues + 4 * N_voxels$$

Goodness-of-fit is assessed per voxel using the coefficient of determination R2 for glucose, lactate, and Glx. A fit is included in downstream analysis if the optimizer reports convergence, or if the mean R2 across channels exceeds 0.5 and the final cost is below 3.0. Bound proximity within 2% of the parameter bounds is flagged diagnostically but does not automatically exclude a fit.

### S3.7 Model Parameter Summary

See **Supp. Table S1**.

## S4. Transcriptome Normalization

### S4.1 Processing of GSE17576 Microarray Data

The transcriptome data for GSE17576 [4] were obtained from the GEO series matrix file for the Affymetrix Mouse Genome 430 2.0 Array. The dataset had already been normalized by the original authors using GC-RMA with the BrainArray Mm_ENTREZG_10 custom CDF, which includes background correction, normalization, and probe summarization at the Entrez gene level. Therefore, no raw microarray normalization was repeated in this workflow.

Because these GC-RMA values were stored in linear-scale, they were transformed using log2(x+1) to place all expression values on a log2 scale for downstream analysis. Probe identifiers from the BrainArray custom CDF were Entrez-based IDs, so they were mapped to current mouse MGI gene symbols using MyGene.info, a gene annotation web service from the BioThings / Su-Wu group at Scripps Research [5]. Since the BrainArray CDF provides one probe set per Entrez gene, no probe-level aggregation was required.

Sample metadata were parsed from the GEO series matrix and annotated by diet and time point. Samples were grouped into three conditions: LFD_day0, LFD_day56, and HFD_day56, with 10 samples in each group. The final normalized GSE17576 expression matrix contained 16,319 genes across 30 samples.

### S4.2 Processing of GSE305719 RNA-Seq Data

GSE305719 [6] was processed from its raw count matrix. The dataset contains RNA-seq profiles across multiple organs and high-fat diet time points. Before normalization, lowly expressed genes were filtered out. Genes were retained only if they had at least 10 reads in at least two samples. The filtered raw count matrix was then normalized using the variance-stabilizing transformation from PyDESeq2 [7]. For this multi-organ dataset, organ was used as the VST (variance-stabilizing transformation) design factor . The VST output was used as a log2-scale, library-size-normalized, variance-stabilized expression matrix. Gene symbols were validated against current mouse gene-symbol annotations using MyGene.info [5]; recognized aliases were updated to current symbols, while unrecognized symbols were kept unchanged.

## S5. Mouse muscle-specific GEM reconstruction

The mouse muscle-specific metabolic model was generated from the iMM1865 genome-scale mouse metabolic reconstruction [8], using normalized microarray-based transcriptomic data from GSE17576 [4]. Samples were grouped by biological condition, and mean expression values were calculated for each gene within each condition to obtain a condition-level expression matrix.

Expression evidence was converted into gene confidence scores for use with CORDA [9]. The expression threshold was defined as the 25th percentile of the per-gene mean expression distribution. Genes with expression above this threshold in all conditions were assigned high confidence, genes above the threshold in only a subset of conditions were assigned medium confidence, and genes below the threshold in all conditions were assigned low confidence. Genes not represented in the transcriptomic dataset were treated as not considered.

Because the expression matrix used gene symbols whereas iMM1865 uses Entrez identifiers, gene symbols were mapped to Entrez IDs using MyGene.info. The resulting gene confidence scores were propagated to reactions using the gene-protein-reaction rules in iMM1865. For reactions associated with alternative isoenzymes, the highest gene confidence score was used. For reactions requiring multi-subunit enzyme complexes, the lowest confidence score among the required genes was used. Reactions without gene associations were assigned a neutral confidence score.

Reaction confidence scores were then provided to CORDA to reconstruct a skeletal muscle-specific model. CORDA retained reactions with strong transcriptomic support where feasible, removed reactions with weak support when they were not required for network consistency, and included supporting reactions needed to maintain a functional metabolic network. The reconstruction was performed using the configured CORDA parameters, with Gurobi as the preferred solver and GLPK as a fallback.

After reconstruction, the model was configured for muscle-oriented FBA. The mitochondrial ATP synthase reaction (ATPS4mi) was set as the objective to represent ATP production capacity. The biomass reaction (BIOMASS_reaction) was constrained to carry at least 80% of its maximum feasible flux, preserving a maintenance requirement during ATP maximization. Both ATPS4mi and BIOMASS_reaction were protected during CORDA reconstruction to ensure their retention in the final model.

The reconstructed model was validated by testing flux feasibility before and after objective reconfiguration. Retention of high-confidence reactions was checked, and selected muscle-relevant reactions, including glycolysis, lactate metabolism, pyruvate oxidation, citrate synthase, fatty acid oxidation, ATP synthase, and biomass maintenance, were inspected. The final muscle-specific model contained 7,653 reactions, 4,913 metabolites, and 1,745 genes.

## S6. Simulating a GEM Constrained by DMI-Derived Reaction Rates

## S6.1 Model Preparation and Exchange-Bound Policy

All simulations were performed using the reconstructed mouse skeletal muscle-specific COBRA model described in **Supp. Text S5**. Before simulation, the model was checked and repaired to ensure that required boundary reactions were present, including the ATP hydrolysis demand reaction DM_atp_c_ and the beta-hydroxybutyrate exchange reaction EX_bhb_e. Known problematic or audited reactions identified were locked to avoid biologically implausible flux routes. These audit-locked reactions were r0191, r0509, r0915, CSNATm, and HMGCOAS, each constrained to zero flux. These locks remove known stoichiometric shortcut or futile-cycle routes, including an alternative UTP-dependent PFK reaction, duplicate succinate oxidation/transport reactions, and an acetyl-carnitine/HMG-CoA artifact cycle.

The model was simulated under a defined nutrient medium. Most boundary reactions were reset to secretion-only by default, whereas selected nutrient exchanges were opened for uptake. The uptake-permitted reactions were glucose, oxygen, hexadecanoate, beta-hydroxybutyrate, and water, with lower bounds:

EX_glc__D_e = -0.11
EX_o2_e = -2.00
EX_hdca_e = -0.08
EX_bhb_e = -0.01
EX_h2o_e = -1000.0

all in mmol $gDW^{-1}$ $h^{-1}$. The biomass reaction lower bound was relaxed to zero in this simulation stage so that the imposed ATP demand and DMI-derived flux constraints could define the feasible region without forcing additional biomass synthesis.

## S6.2 Fixed ATP Demand

To represent a fixed energetic requirement, the ATP hydrolysis demand reaction DM_atp_c_ was constrained to:

$$\text{DM_atp_c_} = 3.5 \text{ mmol gDW}^{-1} \text{ h}^{-1}$$

in all DMI-constrained simulations. This value was selected from the previous calibration of the reference medium, where the maximum attainable ATP demand was approximately 3.56 mmol $gDW^{-1}$ $h^{-1}$ and remained feasible when lactate secretion was additionally imposed. Thus, 3.5 mmol $gDW^{-1}$ $h^{-1}$ was used as a conservative near-maximal ATP requirement that preserved feasibility across all replicate conditions.

## S6.3 DMI Input Data and Replicate Definition

Our workflow constructed one DMI-constrained feasible region per mouse–tissue–week-9 replicate. Thus, each condition row corresponded to a single mouse, one hindlimb muscle tissue, one group label, and week 9. This design preserved mouse-level biological variation rather than averaging mice before GEM simulation. For each tissue replicate, the seed voxel, defined as voxel index 0 in the DMI fitting output, was used as the MRI-derived target source. Other fitted C6 quantities were retained for quality control and audit tables.

## S6.4 DMI-to-GEM Mapping

We imposed two direct GEM constraints:

MR_glc -> HEX1
V_ox -> PDHm

MR_glc was interpreted as a glucose-processing anchor and mapped to the hexokinase reaction HEX1. V_ox was interpreted as glucose-label-derived oxidative pyruvate entry and mapped to mitochondrial pyruvate dehydrogenase, PDHm.

### S6.5 Scaling of DMI Targets to GEM Flux Units

The DMI kinetic model and GEM use different units. The DMI model reports fitted rates in DMI-derived concentration/time units, whereas GEM fluxes are expressed in mmol $gDW^{-1}$ $h^{-1}$. Therefore, a single global MRI-to-FBA scale factor was calculated using the week-9 control reference set.

The average week-9 control MR_glc across the three skeletal muscle tissues was scaled to a reference HEX1 target of:

$$\text{REFERENCE_HEX1_TARGET} = 0.10 \text{ mmol } gDW^{-1} h^{-1}$$

The scale factor was therefore:

$$\text{MRI_TO_FBA_SCALE_FACTOR} = 0.10 / \text{mean}(\text{MR_glc_reference})$$

The same scale factor was then applied to all replicate-level DMI targets. This preserved relative differences among mice, tissues, and groups while placing the DMI-derived anchors into the GEM flux scale.

### S6.6 Soft Target Intervals

DMI-derived fluxes were not imposed as exact equality constraints. Instead, the workflow used soft target intervals. For each selected DMI-derived anchor, the scaled target was converted into an allowed lower and upper interval.

For MR_glc, the default relative tolerance was ±20%:

$$\text{lower_MR_glc} = \text{target_MR_glc} * (1 - 0.20)$$
$$\text{upper_MR_glc} = \text{target_MR_glc} * (1 + 0.20)$$

For V_ox, the default relative tolerance was ±50%:

$$\text{lower_V_ox} = \text{target_V_ox} * (1 - 0.50)$$
$$\text{upper_V_ox} = \text{target_V_ox} * (1 + 0.50)$$

If valid uncertainty-derived intervals were available, these could be used instead. In the current per-replicate workflow, uncertainty was often unavailable at the single seed-voxel target level, so the configured relative tolerances were used by default.

### S6.7 Weighted L1 Mismatch Penalty for Constraint Violations

For each replicate, soft range constraints were added for the selected DMI-derived GEM anchors. For a simulated flux v and an allowed interval [L, U], no penalty was applied if:

$$L <= v <= U$$

If v fell below L, a lower mismatch variable was introduced. If v exceeded U, an upper mismatch variable was introduced. The penalty for each anchor was proportional to the weighted sum of these mismatch variables:

$$\text{penalty_i} = \text{weight_i} / \text{sigma_i} * (\text{mismatch_low_i} + \text{mismatch_high_i})$$

The total DMI mismatch penalty was:

$$P = \text{sum_i penalty_i}$$

where i indexes the selected constraints. The default penalty weights were 1.0 for MR_glc and 0.75 for V_ox. A sigma floor was used to prevent unrealistically small uncertainty estimates from dominating the optimization.

### S6.8 Per-Replicate Feasible Region Construction

For each mouse–tissue–week-9 replicate, the model was copied or freshly loaded, depending on the parallel backend. The simulation then proceeded as follows:

1. Apply the defined nutrient exchange-bound policy.
2. Apply the audited reaction locks.
3. Fix ATP demand with DM_atp_c_ = 3.5.
4. Add soft range constraints for the selected DMI-derived anchors, MR_glc/HEX1 and V_ox/PDHm.
5. Minimize the weighted L1 DMI mismatch penalty P.
6. Convert the optimal penalty P* into a relaxed admissible penalty budget.
7. Constrain the model to remain within that penalty budget, defining the final feasible region for that replicate.

The relaxed penalty budget was computed using the shared settings:

$$\text{P_max} = P^* * (1 + 0.50) + 0.1$$

The final model was therefore not restricted to a single optimal flux distribution. Instead, it represented all flux states that satisfied stoichiometric constraints, nutrient bounds, ATP demand, audit locks, and the relaxed DMI penalty budget.

### S6.9 Flux Variability Analysis and Flux Sampling

Flux variability analysis was performed over the final feasible region for each replicate to compute the minimum and maximum feasible flux for each reaction. Flux sampling was then performed using COBRApy sampling. The sampler was selected automatically from the available COBRApy samplers, using ACHR or OptGP depending on the runtime environment.

For each replicate, the workflow sampled:

$$SAMPLE_COUNT = 50$$

feasible flux vectors with:

$$SAMPLE_THIN = 100$$

The random seed was offset by replicate index, allowing reproducible but distinct sampling for each mouse–tissue condition. The sampling distributions were interpreted as feasible model variation conditional on the imposed DMI-derived constraints, not as independent biological replicates.

### S6.10 Subsystem Activity Scores

A curated reaction-to-subsystem mapping was loaded and filtered to include only reactions present in the muscle-specific model. Reaction-level flux samples were then aggregated into subsystem activity scores.

For each subsystem and each sampled flux vector, subsystem activity was defined as the mean absolute flux across all reactions assigned to that subsystem:

$$A_subsystem = mean(|v_j| \text{ for reactions } j \text{ in subsystem})$$

This produced one subsystem activity score per sampled flux vector. Repeating this across all samples yielded subsystem activity distributions for each mouse, tissue, and group. These distributions were summarized by mean, standard deviation, quantiles, and interquartile ranges, and were used to compare DMI-constrained feasible metabolic activity across tissues and diet conditions.

For exploratory diet-group comparisons, each mouse–tissue replicate was first summarized by the median subsystem activity across its 50 sampled flux vectors. Exact permutation tests were then applied to these replicate-level summaries, so that individual flux samples were not treated as independent biological observations.

## S7. Pathway Activity Scoring and Transcriptomic Comparison

### S7.1 Subsystem-level flux activity

To summarize FBA sampling results at a pathway scale, reactions in the muscle-specific GEM were grouped according to their curated metabolic subsystem annotations. Subsystems are like pathways defined for a GEM and represent sets of reactions participating in related metabolic functions, such as glycolysis, lactate metabolism, pyruvate oxidation, TCA-cycle metabolism, oxidative phosphorylation, fatty-acid uptake and β-oxidation, ketone metabolism, glutamate/α-ketoglutarate metabolism, redox shuttles, and creatine phosphate buffering.

For each sampled flux vector, subsystem activity was calculated as the mean absolute flux across all reactions assigned to a given subsystem. This produced one activity score per subsystem for each feasible sampled metabolic state. Repeating this calculation across all flux samples yielded subsystem activity distributions for each muscle, diet condition, and mouse replicate. These distributions were used to assess how DMI-derived constraints altered the feasible metabolic solution space and whether constrained subsystem activity differed between control and HFD conditions.

To quantify constraint-induced narrowing or expansion of the feasible space, we compared the interquartile range of subsystem activity under DMI-constrained sampling with the corresponding interquartile range in the generic unconstrained reference model:

$$\log2(IQR_{\text{DMI-constrained}} / IQR_{\text{reference}})$$

Negative values indicate that DMI-derived constraints narrowed the feasible activity range of a subsystem, whereas positive values indicate broader feasible activity after applying the constraints. These subsystem activity scores summarize model-feasible flux variation and should not be interpreted as direct measurements of pathway activity.

## S7.2 Transcriptomic gene-module activity

To provide independent biological context for model-predicted subsystem changes, public skeletal-muscle transcriptomic datasets (GSE17576 and GSE305719 [4,6]) were mapped onto the same subsystem structure used for the FBA analysis. Each model reaction is associated with one or more genes through gene-protein-reaction rules. For each metabolic subsystem, all genes associated with reactions in that subsystem were collected to form a subsystem gene module. Duplicate genes arising from multiple reactions within the same subsystem were counted once.

Gene-module activity was summarized separately for each subsystem and transcriptomic comparison. For differential-expression analyses, subsystem-level transcriptomic activity was calculated from the distribution of gene-level $\log_2$ fold changes among genes in the module. The resulting score represents whether genes associated with a given metabolic subsystem tended, on average, to increase or decrease in expression in the disease or diet condition. Enrichment or rank-based statistics were also used to assess whether the subsystem genes showed coordinated transcriptional shifts relative to the background gene set.

These transcriptomic subsystem scores were used only as contextual comparison with the DMI-constrained FBA predictions. Because the transcriptomic datasets were independent of the DMI cohort and not matched by animal, tissue sampling, or experimental protocol, agreement between FBA and RNA was interpreted as supportive biological consistency, whereas disagreement was not treated as direct falsification of the model.

# Supplementary Figures

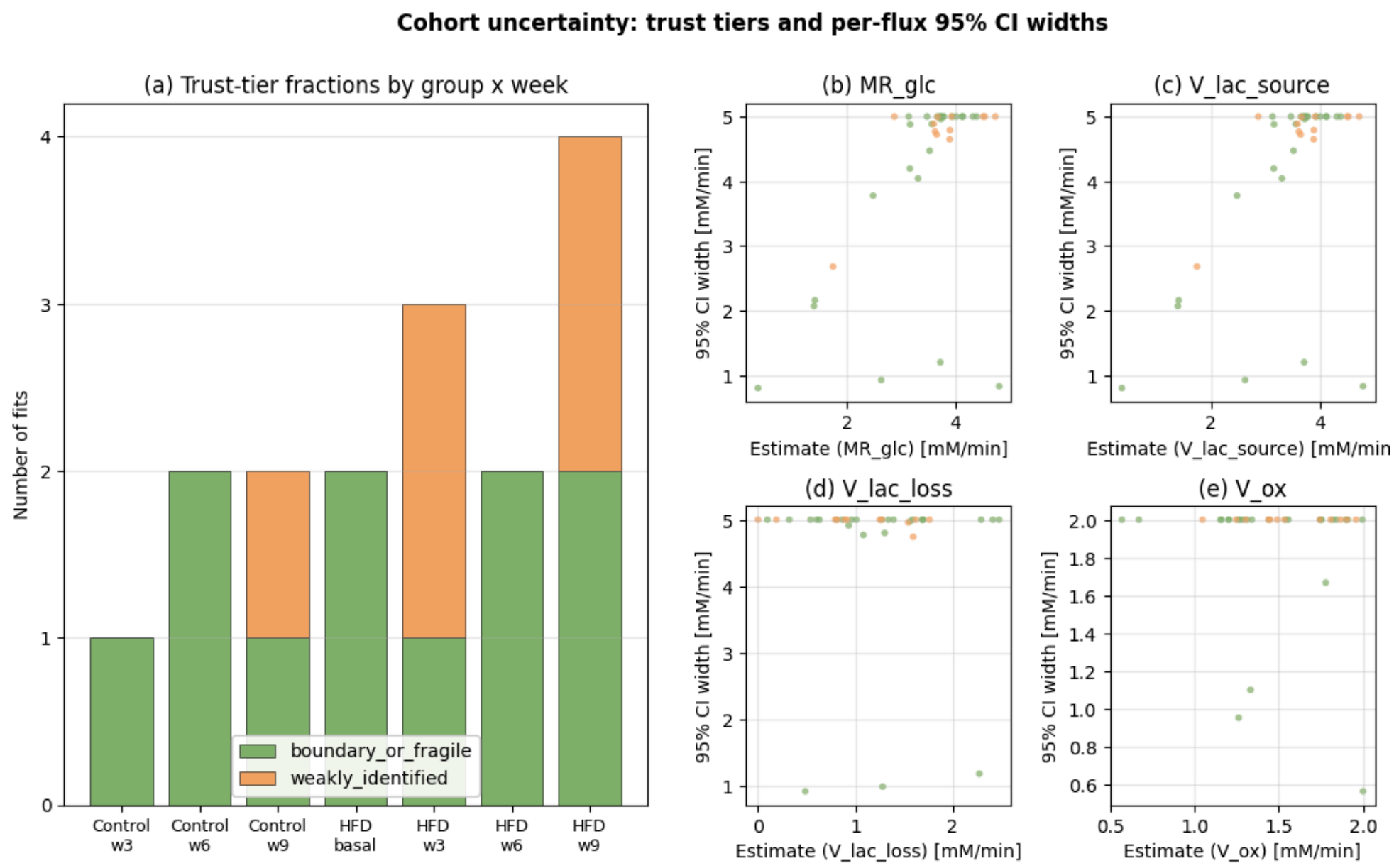

**Figure S1. Cohort uncertainty: trust tiers and per-flux 95% confidence-interval widths.** (a) Stacked bars showing the number of per-mouse fits in each trust tier, broken down by group (Control, HFD) and week. Tiers are assigned from local covariance and profile diagnostics: boundary_or_fragile (green)—fits that are identifiable but constrained by a parameter bound or show mild numerical fragility; weakly_identified (orange) — fits with high parameter correlation or ill-conditioned covariance. (b–e) Scatter of the draw-based 95% CI width against the point estimate for the four cohort-level fluxes carrying Gauss-Newton covariance-based uncertainty propagation draws: glucose metabolic rate MR_glc, lactate efflux V_lac, oxidative flux V_ox (all in mM·min$^{-1}$), and the lactate fraction r_lac (dimensionless). Each marker is one mouse–week fit, coloured by its trust tier as in (a). Larger CI widths at comparable estimates indicate fits whose flux is less tightly constrained by the data.

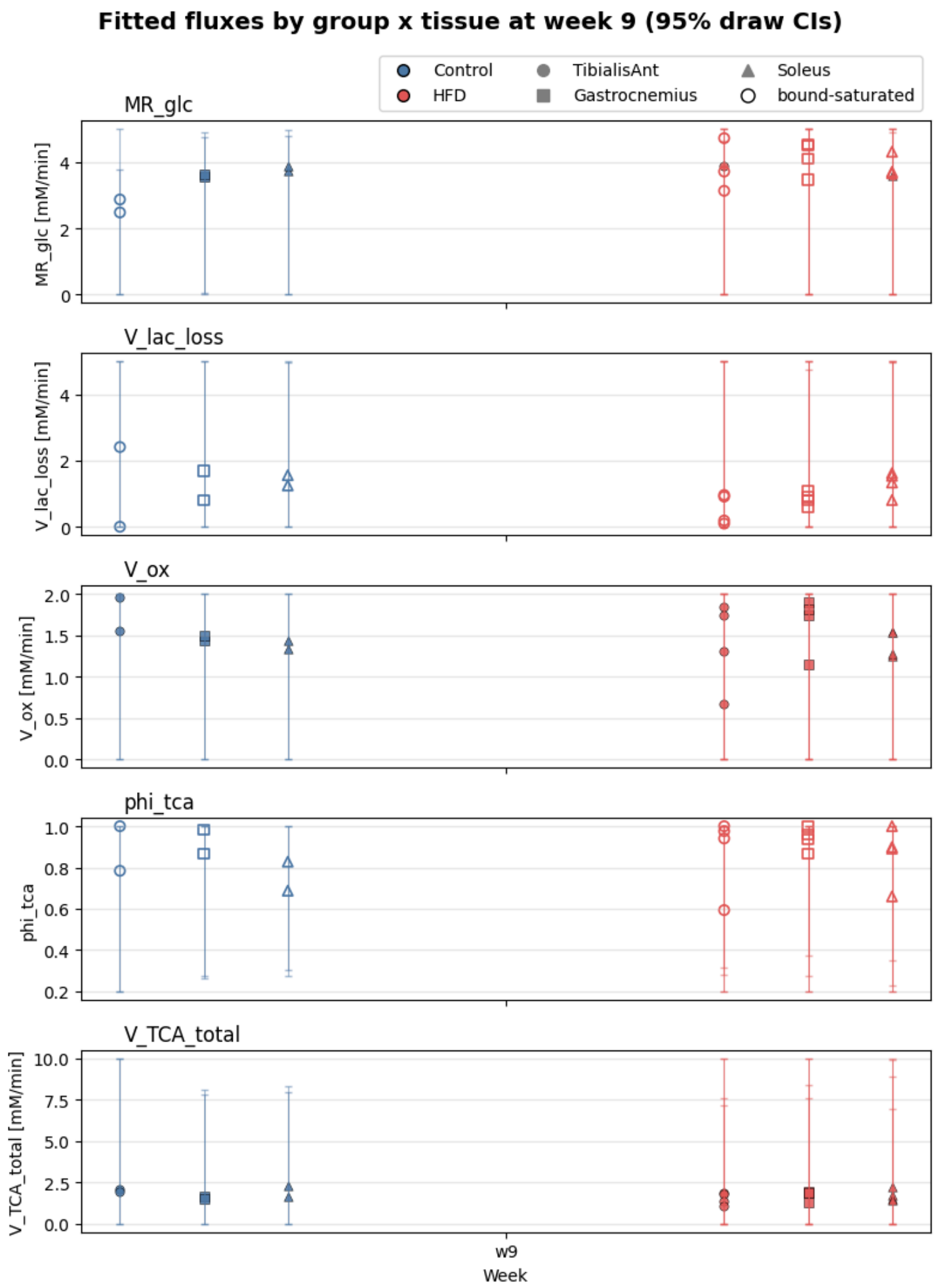

**Figure S2. Fitted DMI-derived fluxes across groups and hindlimb muscles at week 9.** Per-mouse point estimates (filled markers) and 95% confidence intervals from Gauss-Newton covariance-based uncertainty propagation draws for the four parameters that subsequently constrain the FBA model: glucose metabolic rate MR_glc, lactate efflux V_lac_loss, oxidative flux V_ox (all in mM·min$^{-1}$), and the glucose-derived fraction of effective TCA/Glx turnover $\varphi_{TCA}$ (dimensionless). The x-axis is separated by group—Control (blue) vs. HFD (red)—and by tissue, with marker shape encoding muscle (circle = Tibialis anterior, square = Gastrocnemius, triangle = Soleus). Open markers denote fits whose draw-based CI saturates a parameter bound on at least one side, indicating that the interval is bound-limited rather than data-limited. Error bars are omitted where draws are not available for a given flux.

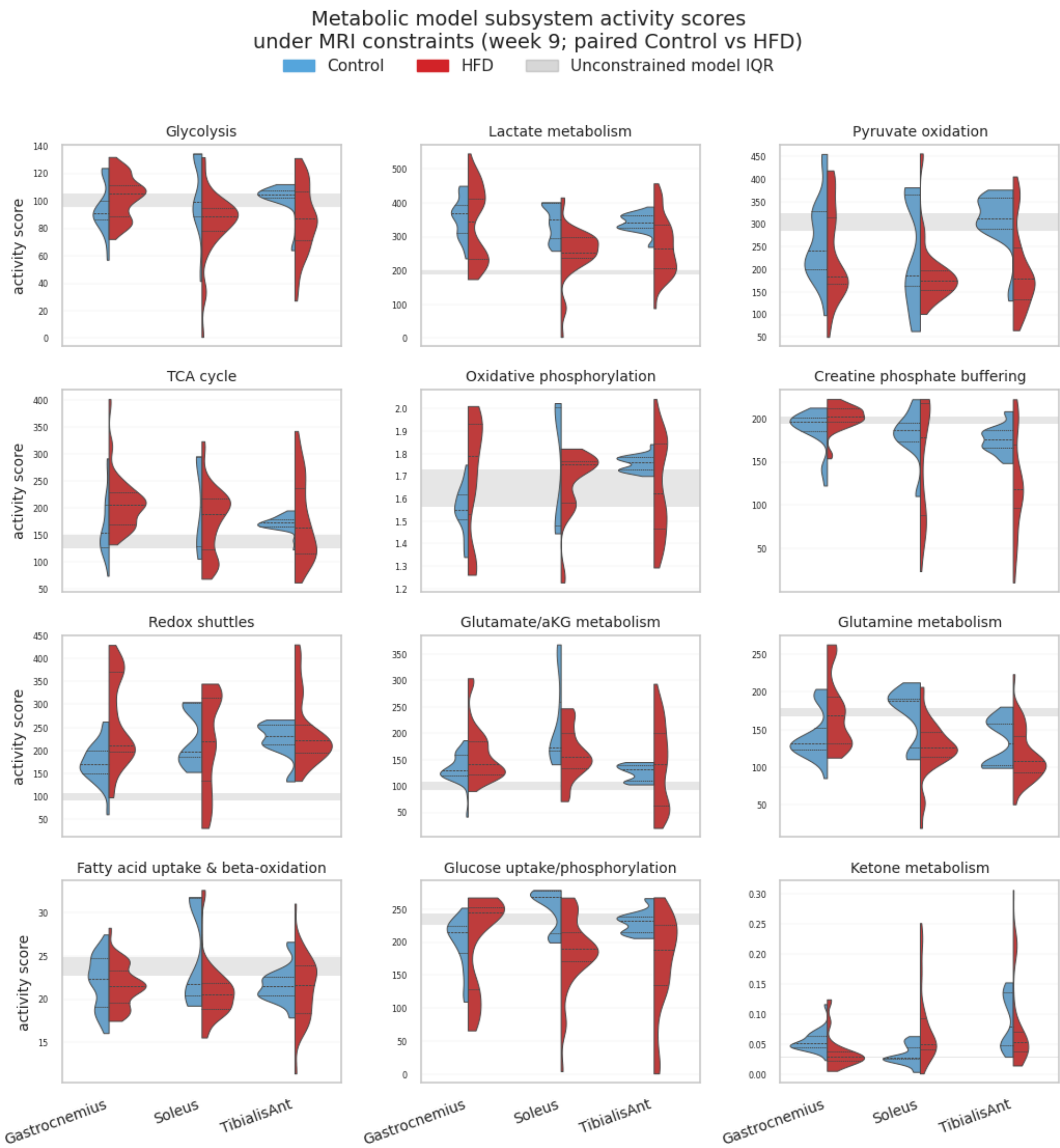

**Figure S3. Subsystem activity distributions from DMI-constrained FBA sampling.** For each muscle subsystem, split violins show sampled subsystem activity scores from DMI-constrained flux sampling for week-9 control and HFD mouse–tissue replicates across Gastrocnemius, Soleus, and Tibialis anterior. Each mouse–tissue replicate contributed 50 feasible flux samples. Grey bands indicate the corresponding interquartile range from the generic unconstrained/reference model. Distributions reflect feasible-space variation under the imposed DMI constraints and should not be interpreted as independent biological replicates.

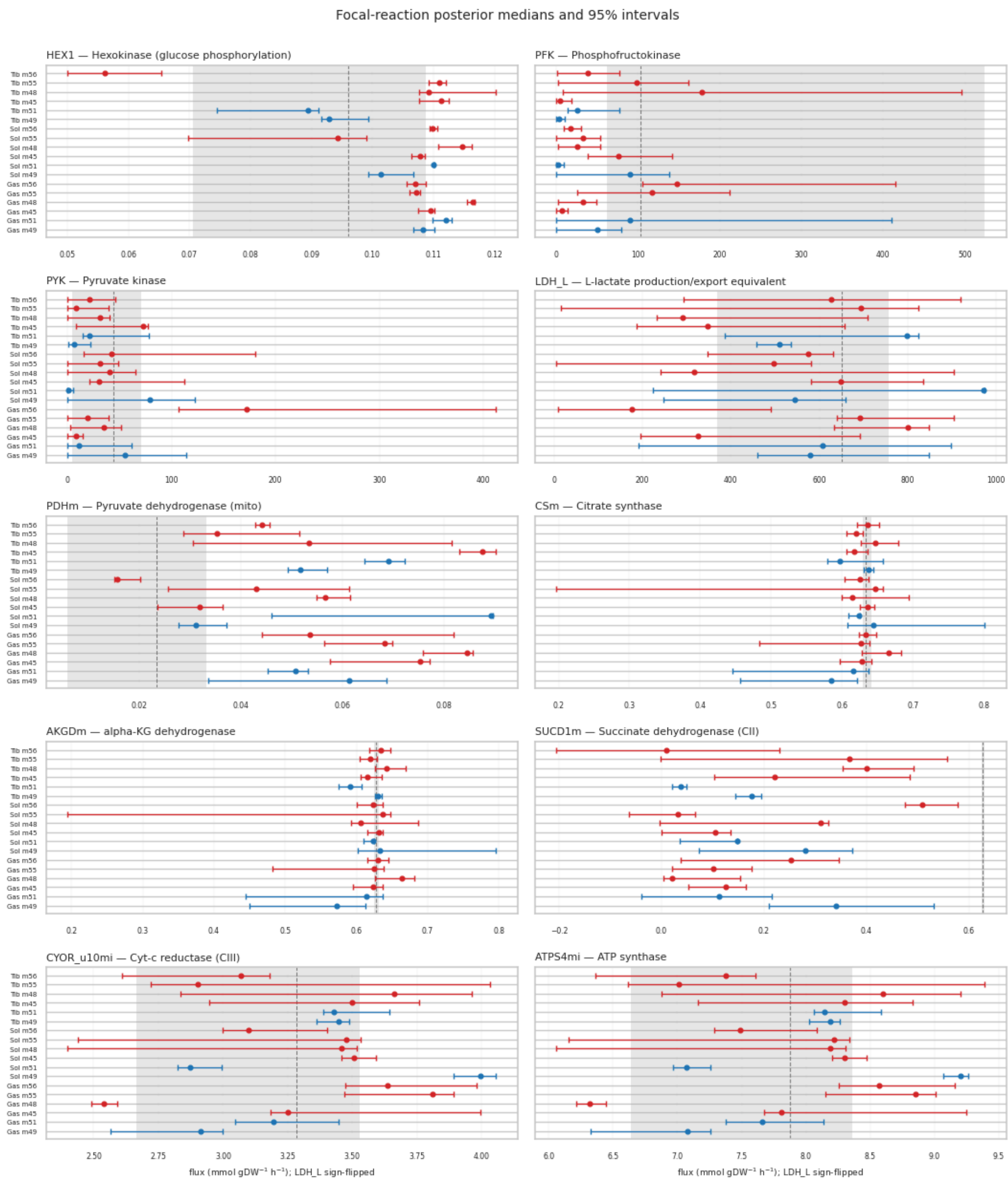

**Figure S4. Focal reaction flux summaries from the DMI-constrained FBA sampling.** Selected reactions span the glucose-to-mitochondrial metabolism axis, including glucose phosphorylation, glycolysis, lactate production/export, pyruvate oxidation, TCA-cycle reactions, electron transport, and ATP synthesis. Each colored point shows the posterior median for one mouse-tissue replicate at week 9 under Control or HFD conditions; horizontal intervals show the 2.5% to 97.5% quantiles across the 50 feasible-region samples for that replicate. The grey band and dashed line show the 95% interval and median reference. LDH_L is sign-flipped so positive values correspond to the lactate-production/export interpretation used elsewhere in the project. These reaction-level shifts are model-implied and should be interpreted as constrained feasible-space changes rather than direct biochemical flux measurements.

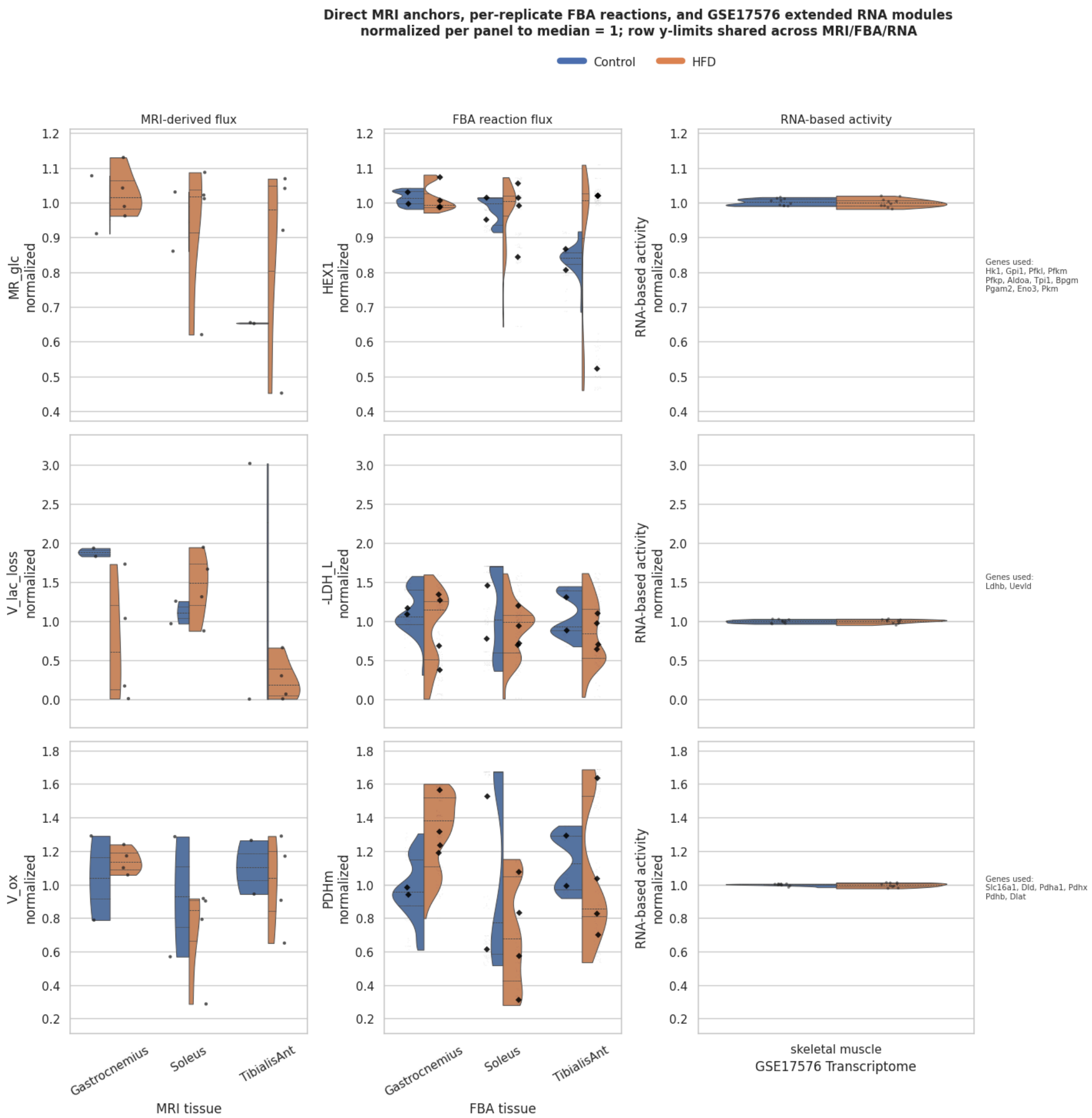

**Figure S5. Direct comparison of DMI-derived fluxes, DMI-constrained FBA reaction fluxes, and transcriptomic reaction-gene modules in GSE17576.** For each DMI-linked metabolic quantity, the left panels show the week-9 per-replicate DMI-derived flux estimates, the middle panels show the corresponding reaction flux distributions from DMI-constrained FBA sampling, and the right panels show expression changes of reaction-associated gene modules in the independent GSE17576 skeletal-muscle transcriptomic dataset [4]. DMI quantities were mapped to model reactions as follows to serve as constraints: MR_glc to HEX1 and V_ox to PDHm; V_lac,loss is mapped to LDH_L with sign adjustment for lactate-producing direction, serving as a diagnosis rather than a constraint. Points or distributions represent mouse-level DMI/FBA replicates where applicable; transcriptomic panels summarize genes linked to the corresponding model reactions through gene-protein-reaction annotations. Note that the RNA dataset is independent of the DMI cohort.

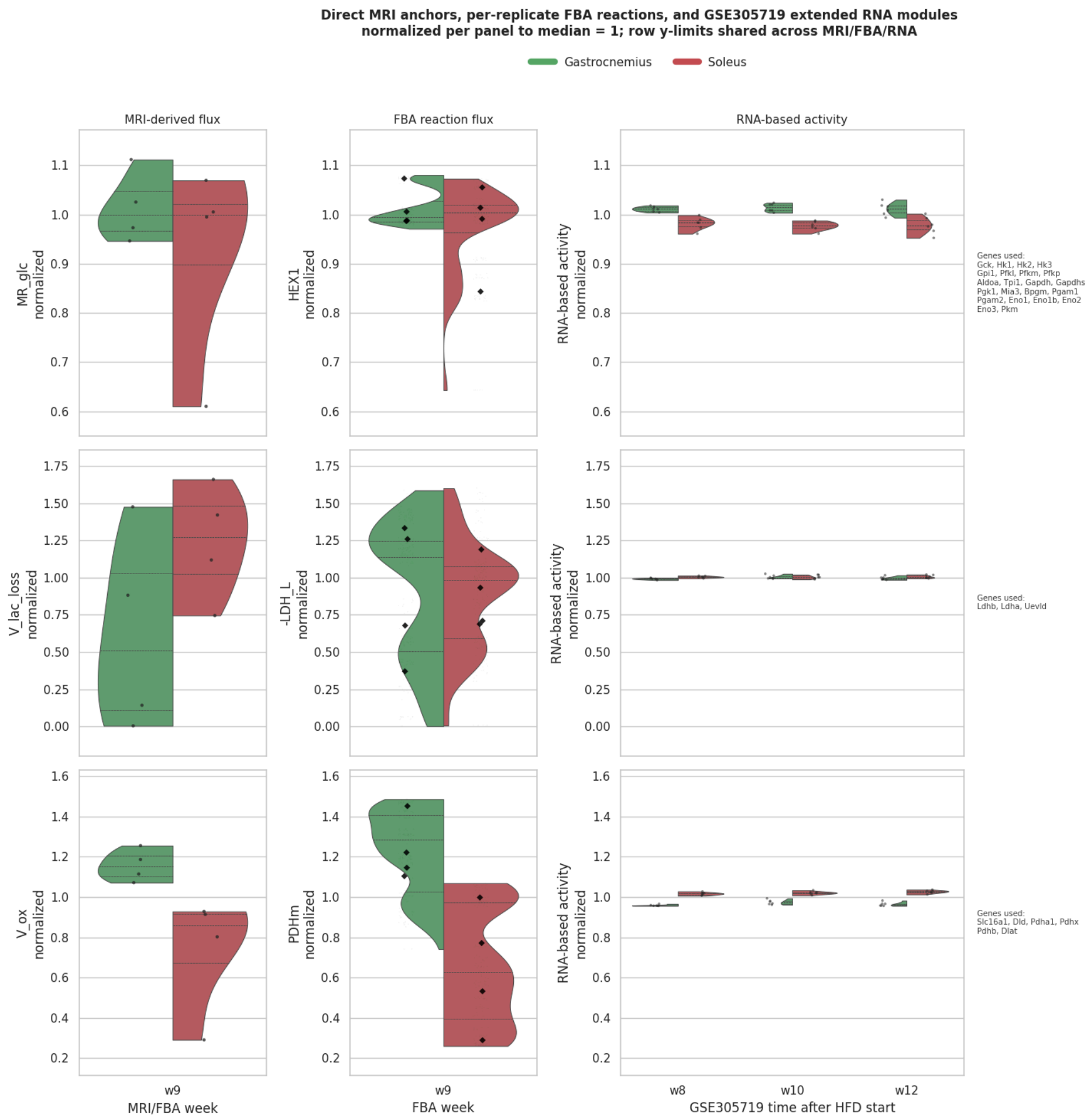

**Figure S6. Direct comparison of DMI-derived fluxes, DMI-constrained FBA reaction fluxes, and transcriptomic reaction-gene modules in GSE305719.** Per-replicate DMI-derived flux estimates and their corresponding DMI-constrained FBA reaction fluxes are compared with reaction-gene module expression in the independent GSE305719 skeletal-muscle transcriptomic dataset [6]. The figure links each DMI-derived constraint to its associated model reaction: glucose metabolism (MR_glc/HEX1) and pyruvate oxidation (V_ox/PDHm); for lactate production, the V_lac,loss/LDH_L comparison (sign-adjusted) serves as diagnosis and not a constraints. The DMI and FBA panels summarize week-9 muscle-specific model inputs and sampled flux outputs, whereas the transcriptomic panels summarize expression behavior of genes associated with the same reactions. These panels are intended to visualize cross-modal consistency and heterogeneity across DMI fluxes, model-predicted fluxes, and gene expression, not to provide one-to-one validation across matched biological samples.

# Supplementary Tables

### Table S1. Model parameters, fixed values, and optimization bounds.

| Symbol | Description | Value / Bounds | Status | Source |
|---|---|---|---|---|
| f_gly | Glycolytic label-loss correction factor | 0.157 | Fixed | Li et al. 2025 [2] |
| f_tca | TCA cycle label-loss correction factor | 0.385 | Fixed | Li et al. 2025 [2] |

| Symbol | Description | Value / Bounds | Status | Source |
|---|---|---|---|---|
| kie_gly | Kinetic isotope effect — glycolysis | 1.042 | Fixed | de Graaf et al. 2021 [10] |
| kie_tca | Kinetic isotope effect — TCA cycle | 1.035 | Fixed | de Graaf et al. 2021 [10] |
| K_T | GLUT4 Michaelis constant | 5.8 mM | Fixed | Mathy et al. 2026 [3] |
| Glc_p,total | Total plasma glucose (unlabeled baseline) | 8.0 mM | Fixed | Assumed; fasted lean mouse |
| Glc_cell | Intracellular glucose pool size | 8.0 mM | Fixed | Lumped compartment assumption |
| [Lac] | Lactate/pyruvate pool size | 1.5 mM | Fixed | Assumed |
| [aKG] | α-Ketoglutarate pool size | 0.10 mM | Fixed | Assumed |
| [Glx] | Glutamate + glutamine pool size | 10.0 mM | Fixed | Assumed |
| T_max | Maximum glucose transport rate | $\eta \cdot MR_{glc}$ | Derived | — |
| V_x | α-KG ↔ Glx exchange flux | [0.01, 10.0] μmol/g/min | Free (tissue) | — |
| f_v | Blood volume fraction (partial volume) | [0.0, 0.15] | Free (tissue) | — |
| η | Transport capacity multiplier, $T_{max} = \eta \cdot MR_{glc}$ | [1.0, 15.0] | Free (tissue) | — |
| MR_glc | Glucose metabolic rate | [0.001, 5.0] μmol/g/min | Free (voxel) | — |
| V_lac,loss | Apparent Lac/Pyr label clearance or dilution | [0.0, 5.0] μmol/g/min | Free (voxel) | — |
| V_ox | Glucose-label-derived oxidative entry | [0.0, 2.0] μmol/g/min | Free (voxel) | — |
| φ_TCA | Glucose-derived fraction of effective TCA/Glx turnover | [0.2, 1.0] | Free (voxel) | — |
| V_TCA,total | Total effective TCA/Glx turnover | V_ox/φ_TCA | Derived | — |
| V_dil,TCA | Apparent non-glucose/dilution component of TCA/Glx turnover | V_TCA,total−V_ox | Derived | — |
| $a_1$ | Plasma input amplitude | [10, 60] mM | Free (global) | — |
| Δt_delay | Bolus arrival delay | [0, 12] min | Free (global) | — |
| τ_rise | Plasma rise time constant | [0.2, 12] min | Free (global) | — |
| w | Fast-decay weight | [0.05, 0.95] | Free (global) | — |
| k_p,fast | Fast plasma decay rate | Derived from k_p,slow + k_p,gap | Free (global) | — |

| Symbol | Description | Value / Bounds | Status | Source |
|---|---|---|---|---|
| k_p,slow | Slow plasma decay rate | [0.001, 0.030] $min^{-1}$ | Free (global) | — |
| k_p,gap | Difference k_p,fast − k_p,slow | [0.002, 0.110] $min^{-1}$ | Free (global) | — |

Abbreviations: TA = tibialis anterior; KIE = kinetic isotope effect; DE = differential evolution; L-BFGS-B = limited-memory Broyden–Fletcher–Goldfarb–Shanno with box constraints.

**Table S2. Mapping of DMI-derived metabolic flux estimates to GEM constraints.** In the default C6-aware mode, only MR_glc and V_ox were used as direct GEM constraints. Lactate and TCA/dilution-related quantities were retained for audit and diagnostic interpretation but were not imposed as direct reaction constraints.

| Quantity | Biological description | FBA reaction / expression | FBA interpretation | Constraint status |
|---|---|---|---|---|
| MR_glc | DMI-derived glucose consumption / glucose-label entry | HEX1 | Hexokinase flux; intracellular glucose phosphorylation proxy | Used as soft range constraint |
| V_ox | Glucose-label-derived oxidative pyruvate entry | PDHm | Mitochondrial pyruvate dehydrogenase flux | Used as soft range constraint |
| V_lac_loss | Labeled Lac/Pyr-pool clearance or dilution | Not directly mapped | Not equivalent to LDH flux in FBA model | Diagnostic only |
| V_out | Compatibility alias related to lactate-label clearance | Not directly mapped | Not used as lactate export constraint | Diagnostic only |
| V_lac | Compatibility alias retained from older outputs | Not mapped to LDH_L | Guardrail prevents inappropriate LDH constraint | Diagnostic only |
| V_TCA_total | Derived total effective TCA/Glx turnover | Not directly mapped | Model-derived turnover quantity, not direct total TCA flux | Diagnostic only |
| V_dil_tca | Derived apparent dilution/non-glucose component | Not directly mapped | Indicates dilution contribution in fit | Diagnostic only |
| phi_tca | Glucose-derived fraction of effective TCA/Glx turnover | Not directly mapped | Fractional dilution diagnostic | Diagnostic only |

**Table S3. Technical implementation of precautionary modeling strategies used in the DMI-to-GEM workflow.** This table expands Table 2 by listing the specific modeling choices, implementation details, interpretation rules, and suggested sensitivity analyses associated with each uncertainty source. These strategies were used to prevent weakly identified DMI-derived quantities from being interpreted as direct biochemical pathway fluxes and to ensure that DMI-constrained FBA/FVA/sampling results were treated as model-conditioned feasible-space predictions.

| **Uncertainty source** | **Implementation in current workflow** | **Interpretation rule** | **Suggested future sensitivity / validation** |
|---|---|---|---|
| Unequal DMI signal reliability across metabolites | Glc* fit quality was treated as more reliable than Lac* and Glx*. MR_glc and V_ox were used as direct GEM inputs; Lac/Glx-dependent quantities were retained diagnostically. | Biological conclusions should rely mainly on glucose-derived constraints, not Lac/Glx-derived absolute fluxes. | Compare conclusions across spectral-processing pipelines; report fit-quality thresholds and metabolite-specific $R^2$. |
| Mapping DMI quantities to GEM reactions | MR_glc → HEX1; V_ox → PDHm. V_lac,loss, V_TCA,total, V_dil,TCA, and φ_TCA were not directly imposed as GEM constraints. | HEX1 and PDHm are DMI-informed anchors, not direct measurements of total glycolytic or TCA-cycle flux. | Test MR_glc-only, V_ox-only, and MR_glc+V_ox models as ablations. |
| DMI-to-GEM unit conversion | A single global MRI-to-FBA scale factor was derived from the week-9 control MR_glc reference set and applied to all replicates. | Relative differences among mouse, tissue, and diet groups are more interpretable than absolute GEM flux magnitudes. | Vary the scale factor, for example 0.5×, 1×, and 2×, and test whether subsystem-level conclusions remain directionally stable. |
| Constraint uncertainty | MR_glc and V_ox were imposed as soft target intervals rather than exact equality constraints. | Constraint satisfaction should be interpreted as compatibility with DMI-derived estimates, not exact reproduction of measured fluxes. | Test narrower and wider soft intervals, for example MR_glc ±10–30% and V_ox ±25–75%. |
| GEM feasibility under DMI constraints | Weighted L1 mismatch penalties allowed deviations outside target intervals when required for feasibility. The relaxed penalty budget defined the final feasible region. | A feasible solution indicates stoichiometric compatibility under the chosen assumptions, not proof of biological correctness. | Report constraint violation penalties and identify replicates requiring large relaxation. |
| ATP demand assumption | ATP hydrolysis demand was fixed through DM_atp_c_. | Predicted subsystem shifts are conditional on the chosen energetic demand. | Repeat simulations with lower and higher ATP demand values. |
| Exchange-bound assumptions | Uptake was allowed for selected nutrients such as glucose, oxygen, hexadecanoate, beta-hydroxybutyrate, and water; other exchanges were secretion-only by default. | Feasible-space changes depend on the assumed muscle medium. | Test alternative nutrient-bound policies, especially fatty-acid and oxygen uptake. |
| FBA sampling interpretation | Each mouse–tissue replicate generated sampled feasible flux vectors. Samples were interpreted as feasible model variation, not biological replicates. | Do not use individual sampled flux vectors as independent statistical observations. | Summarize first at replicate level, then compare group-level medians or directions. |
| Low sample size | Results were summarized at mouse/tissue replicate level and interpreted descriptively. | HFD–Control patterns are exploratory and hypothesis-generating. | Increase biological sample size or validate selected predictions experimentally. |
| Missing plasma input measurements | Plasma $^2$H-glucose input was fitted jointly across tissues for each mouse-week rather than directly measured. | Absolute kinetic parameters may be affected by uncertainty in the input function. | Measure plasma glucose enrichment and plasma lactate enrichment in future experiments. |
| Missing lactate uptake in ODE model | The model includes apparent Lac/Pyr label clearance but not explicit plasma lactate uptake or bidirectional lactate exchange. | $V_{lac,loss}$ should not be interpreted as LDH flux or lactate export alone. | Extend the ODE system with plasma lactate input, lactate uptake, lactate export, and lactate oxidation. |
| Transcriptomic comparison | RNA module scores were computed from independent public datasets, not matched DMI animals. | RNA agreement supports plausibility; RNA disagreement does not invalidate DMI/FBA predictions. | Add matched RNA-seq, qPCR, metabolomics, enzyme assays, or isotope tracing. |

**Table S4. Replicate-level exploratory comparison of DMI-constrained subsystem activity between HFD and control mice.** For each mouse–tissue replicate, subsystem activity was summarized as the median across 50 feasible flux samples, and HFD–Control differences were tested using exact permutation tests on mouse-level summaries. Because each comparison contained only two control and four HFD mouse-level replicates, P values were interpreted as descriptive support for directional trends rather than confirmatory

evidence. Individual flux samples were not treated as independent biological replicates. (Gas: Gastrocnemius; Sol: Soleus; TA: Tibialis anterior)

| Tissue | Comparison | Subsystem | Ctrl median | HFD median | Difference | Direction | Permutation P | Welch P | Cliffs delta | N Ctrl | N HFD |
|---|---|---|---|---|---|---|---|---|---|---|---|
| Gas | HFD vs Ctrl | Creatine phosphate buffering | 1.96E+02 | 2.04E+02 | 8.44E+00 | higher in HFD | 0.400 | 0.185 | 0.500 | 2 | 4 |
| Gas | HFD vs Ctrl | Fatty acid uptake & beta-oxidation | 2.09E+01 | 2.16E+01 | 7.06E-01 | higher in HFD | 0.867 | 0.781 | 0.000 | 2 | 4 |
| Gas | HFD vs Ctrl | Glucose uptake/phosphorylation | 2.03E+02 | 2.47E+02 | 4.40E+01 | higher in HFD | 0.600 | 0.897 | 0.500 | 2 | 4 |
| Gas | HFD vs Ctrl | Glutamate/aKG metabolism | 1.30E+02 | 1.38E+02 | 8.85E+00 | higher in HFD | 0.667 | 0.390 | 0.500 | 2 | 4 |
| Gas | HFD vs Ctrl | Glutamine metabolism | 1.32E+02 | 1.53E+02 | 2.15E+01 | higher in HFD | 0.667 | 0.264 | 0.500 | 2 | 4 |
| Gas | HFD vs Ctrl | Glycolysis | 9.36E+01 | 1.03E+02 | 9.08E+00 | higher in HFD | 0.533 | 0.390 | 0.500 | 2 | 4 |
| Gas | HFD vs Ctrl | Ketone metabolism | 4.93E-02 | 2.86E-02 | -2.07E-02 | lower in HFD | 0.067 | 0.022 | -1.000 | 2 | 4 |
| Gas | HFD vs Ctrl | Lactate metabolism | 3.47E+02 | 3.27E+02 | -2.01E+01 | lower in HFD | 0.867 | 0.831 | 0.000 | 2 | 4 |
| Gas | HFD vs Ctrl | Oxidative phosphorylation | 1.56E+00 | 1.76E+00 | 1.94E-01 | higher in HFD | 0.533 | 0.481 | 0.500 | 2 | 4 |
| Gas | HFD vs Ctrl | Pyruvate oxidation | 2.68E+02 | 1.77E+02 | -9.12E+01 | lower in HFD | 0.467 | 0.658 | -0.500 | 2 | 4 |
| Gas | HFD vs Ctrl | Redox shuttles | 1.68E+02 | 2.09E+02 | 4.06E+01 | higher in HFD | 0.400 | 0.158 | 1.000 | 2 | 4 |
| Gas | HFD vs Ctrl | TCA cycle | 1.48E+02 | 2.08E+02 | 6.00E+01 | higher in HFD | 0.133 | 0.070 | 1.000 | 2 | 4 |
| Sol | HFD vs Ctrl | Creatine phosphate buffering | 1.84E+02 | 1.45E+02 | -3.87E+01 | lower in HFD | 0.867 | 0.390 | 0.000 | 2 | 4 |
| Sol | HFD vs Ctrl | Fatty acid uptake & beta-oxidation | 2.61E+01 | 2.01E+01 | -5.95E+00 | lower in HFD | 0.200 | 0.491 | -0.500 | 2 | 4 |
| Sol | HFD vs Ctrl | Glucose uptake/phosphorylation | 2.46E+02 | 1.87E+02 | -5.92E+01 | lower in HFD | 0.133 | 0.367 | -0.750 | 2 | 4 |
| Sol | HFD vs Ctrl | Glutamate/aKG metabolism | 1.70E+02 | 1.53E+02 | -1.73E+01 | lower in HFD | 0.667 | 0.778 | -0.500 | 2 | 4 |
| Sol | HFD vs Ctrl | Glutamine metabolism | 1.57E+02 | 1.23E+02 | -3.43E+01 | lower in HFD | 0.133 | 0.478 | -0.750 | 2 | 4 |
| Sol | HFD vs Ctrl | Glycolysis | 1.16E+02 | 8.80E+01 | -2.76E+01 | lower in HFD | 0.133 | 0.375 | -0.750 | 2 | 4 |
| Sol | HFD vs Ctrl | Ketone metabolism | 3.47E-02 | 5.22E-02 | 1.76E-02 | higher in HFD | 0.333 | 0.131 | 1.000 | 2 | 4 |
| Sol | HFD vs Ctrl | Lactate metabolism | 3.52E+02 | 2.71E+02 | -8.15E+01 | lower in HFD | 0.067 | 0.275 | -1.000 | 2 | 4 |
| Sol | HFD vs Ctrl | Oxidative phosphorylation | 1.74E+00 | 1.75E+00 | 9.28E-03 | higher in HFD | 1.000 | 0.930 | 0.000 | 2 | 4 |
| Sol | HFD vs Ctrl | Pyruvate oxidation | 2.65E+02 | 1.78E+02 | -8.65E+01 | lower in HFD | 0.267 | 0.529 | -0.250 | 2 | 4 |
| Sol | HFD vs Ctrl | Redox shuttles | 2.45E+02 | 2.63E+02 | 1.80E+01 | higher in HFD | 0.800 | 0.878 | 0.250 | 2 | 4 |
| Sol | HFD vs Ctrl | TCA cycle | 2.12E+02 | 1.97E+02 | -1.52E+01 | lower in HFD | 0.933 | 0.749 | -0.250 | 2 | 4 |
| TA | HFD vs Ctrl | Creatine phosphate buffering | 1.76E+02 | 1.23E+02 | -5.34E+01 | lower in HFD | 0.400 | 0.189 | -0.500 | 2 | 4 |
| TA | HFD vs Ctrl | Fatty acid uptake & beta-oxidation | 2.15E+01 | 2.10E+01 | -4.62E-01 | lower in HFD | 0.933 | 0.672 | 0.000 | 2 | 4 |
| TA | HFD vs Ctrl | Glucose uptake/phosphorylation | 2.26E+02 | 1.92E+02 | -3.40E+01 | lower in HFD | 0.333 | 0.159 | -0.750 | 2 | 4 |
| TA | HFD vs Ctrl | Glutamate/aKG metabolism | 1.25E+02 | 1.41E+02 | 1.65E+01 | higher in HFD | 0.800 | 0.806 | 0.250 | 2 | 4 |
| TA | HFD vs Ctrl | Glutamine metabolism | 1.30E+02 | 1.07E+02 | -2.34E+01 | lower in HFD | 0.667 | 0.682 | -0.250 | 2 | 4 |
| TA | HFD vs Ctrl | Glycolysis | 1.05E+02 | 9.22E+01 | -1.28E+01 | lower in HFD | 0.533 | 0.284 | -0.500 | 2 | 4 |
| TA | HFD vs Ctrl | Ketone metabolism | 9.29E-02 | 5.52E-02 | -3.78E-02 | lower in HFD | 0.933 | 0.982 | 0.000 | 2 | 4 |
| TA | HFD vs Ctrl | Lactate metabolism | 3.44E+02 | 2.61E+02 | -8.30E+01 | lower in HFD | 0.400 | 0.216 | -0.500 | 2 | 4 |
| TA | HFD vs Ctrl | Oxidative phosphorylation | 1.75E+00 | 1.66E+00 | -8.87E-02 | lower in HFD | 0.933 | 0.427 | 0.000 | 2 | 4 |
| TA | HFD vs Ctrl | Pyruvate oxidation | 3.23E+02 | 1.83E+02 | -1.39E+02 | lower in HFD | 0.267 | 0.100 | -0.750 | 2 | 4 |
| TA | HFD vs Ctrl | Redox shuttles | 2.34E+02 | 2.17E+02 | -1.71E+01 | lower in HFD | 0.800 | 0.925 | -0.250 | 2 | 4 |
| TA | HFD vs Ctrl | TCA cycle | 1.72E+02 | 1.60E+02 | -1.17E+01 | lower in HFD | 0.933 | 0.979 | 0.000 | 2 | 4 |

## Supplementary Files

**File S1. Week-9 DMI-derived flux estimates used for FBA constraints.** Mouse-level fitted flux estimates were summarized by muscle and disease state for the week-9 conditions used in the DMI-constrained FBA analysis. Values report the mean, between mouse SD (denotes variability of the mouse-level point estimates within each tissue/state group and is distinct from the wider draw-based uncertainty intervals used to represent parameter-estimation uncertainty.), and median across mouse-tissue fits, with HFD minus Control differences shown as absolute and percent changes. Fluxes include glucose-processing rate (MR_glc), lactate-associated flux (V_lac), oxidative flux proxy (V_ox), glucose-label-derived TCA flux (V_TCA), lactate export-associated flux (V_out), and lactate fraction (r_lac). The table is intended to show the physiological coherence of the central fitted rates, while uncertainty is shown separately in the fitted-flux interval plot.

**supplementary_file_S1**

| Tissue | State | Flux | N mouse tissue | Mean | Between Mouse SD | Median | HFD - Control | HFD - Control [%] |
|---|---|---|---|---|---|---|---|---|
| Gastrocnemius | Control | MR_glc | 2 | 4,378 | 0,519 | 4,378 | 0,162 | 3,701 |
| Gastrocnemius | HFD | MR_glc | 4 | 4,541 | 0,326 | 4,474 | 0,162 | 3,701 |
| Gastrocnemius | Control | V_lac_loss | 2 | 1,905 | 0,076 | 1,905 | -1,161 | -60,972 |
| Gastrocnemius | HFD | V_lac_loss | 4 | 0,743 | 0,812 | 0,608 | -1,161 | -60,972 |
| Gastrocnemius | Control | V_ox | 2 | 1,609 | 0,551 | 1,609 | 0,16 | 9,913 |
| Gastrocnemius | HFD | V_ox | 4 | 1,769 | 0,124 | 1,759 | 0,16 | 9,913 |
| Gastrocnemius | Control | phi_tca | 2 | 0,978 | 0,02 | 0,978 | -0,024 | -2,486 |
| Gastrocnemius | HFD | phi_tca | 4 | 0,954 | 0,043 | 0,967 | -0,024 | -2,486 |
| Gastrocnemius | Control | V_TCA_total | 2 | 1,651 | 0,596 | 1,651 | 0,203 | 12,271 |
| Gastrocnemius | HFD | V_TCA_total | 4 | 1,853 | 0,081 | 1,864 | 0,203 | 12,271 |
| Gastrocnemius | Control | V_dil_tca | 2 | 0,041 | 0,045 | 0,041 | 0,043 | 103,754 |
| Gastrocnemius | HFD | V_dil_tca | 4 | 0,085 | 0,08 | 0,061 | 0,043 | 103,754 |
| Gastrocnemius | Control | V_ox_over_MR_glc | 2 | 0,378 | 0,17 | 0,378 | 0,013 | 3,33 |
| Gastrocnemius | HFD | V_ox_over_MR_glc | 4 | 0,39 | 0,026 | 0,381 | 0,013 | 3,33 |
| Gastrocnemius | Control | V_TCA_total_over_MR_glc | 2 | 0,388 | 0,182 | 0,388 | 0,022 | 5,591 |
| Gastrocnemius | HFD | V_TCA_total_over_MR_glc | 4 | 0,409 | 0,03 | 0,405 | 0,022 | 5,591 |
| Soleus | Control | MR_glc | 2 | 4,165 | 0,53 | 4,165 | -0,046 | -1,093 |
| Soleus | HFD | MR_glc | 4 | 4,119 | 0,936 | 4,478 | -0,046 | -1,093 |
| Soleus | Control | V_lac_loss | 2 | 1,124 | 0,206 | 1,124 | 0,343 | 30,496 |
| Soleus | HFD | V_lac_loss | 4 | 1,467 | 0,468 | 1,506 | 0,343 | 30,496 |
| Soleus | Control | V_ox | 2 | 1,436 | 0,786 | 1,436 | -0,315 | -21,921 |
| Soleus | HFD | V_ox | 4 | 1,122 | 0,46 | 1,312 | -0,315 | -21,921 |
| Soleus | Control | phi_tca | 2 | 0,897 | 0,014 | 0,897 | -0,006 | -0,661 |
| Soleus | HFD | phi_tca | 4 | 0,891 | 0,138 | 0,935 | -0,006 | -0,661 |
| Soleus | Control | V_TCA_total | 2 | 1,595 | 0,851 | 1,595 | -0,375 | -23,512 |
| Soleus | HFD | V_TCA_total | 4 | 1,22 | 0,409 | 1,349 | -0,375 | -23,512 |
| Soleus | Control | V_dil_tca | 2 | 0,159 | 0,065 | 0,159 | -0,06 | -37,907 |
| Soleus | HFD | V_dil_tca | 4 | 0,099 | 0,089 | 0,099 | -0,06 | -37,907 |
| Soleus | Control | V_ox_over_MR_glc | 2 | 0,336 | 0,146 | 0,336 | -0,041 | -12,123 |
| Soleus | HFD | V_ox_over_MR_glc | 4 | 0,295 | 0,169 | 0,284 | -0,041 | -12,123 |
| Soleus | Control | V_TCA_total_over_MR_glc | 2 | 0,373 | 0,157 | 0,373 | -0,051 | -13,678 |
| Soleus | HFD | V_TCA_total_over_MR_glc | 4 | 0,322 | 0,176 | 0,29 | -0,051 | -13,678 |
| Tibialis anterior | Control | MR_glc | 2 | 2,879 | 0,008 | 2,879 | 0,956 | 33,212 |
| Tibialis anterior | HFD | MR_glc | 4 | 3,835 | 1,261 | 4,32 | 0,956 | 33,212 |
| Tibialis anterior | Control | V_lac_loss | 2 | 1,528 | 2,161 | 1,528 | -1,269 | -83,029 |
| Tibialis anterior | HFD | V_lac_loss | 4 | 0,259 | 0,301 | 0,184 | -1,269 | -83,029 |
| Tibialis anterior | Control | V_ox | 2 | 1,709 | 0,351 | 1,709 | -0,154 | -9,036 |
| Tibialis anterior | HFD | V_ox | 4 | 1,555 | 0,441 | 1,607 | -0,154 | -9,036 |
| Tibialis anterior | Control | phi_tca | 2 | 0,929 | 0,1 | 0,929 | -0,021 | -2,227 |
| Tibialis anterior | HFD | phi_tca | 4 | 0,909 | 0,153 | 0,977 | -0,021 | -2,227 |
| Tibialis anterior | Control | V_TCA_total | 2 | 1,829 | 0,181 | 1,829 | -0,139 | -7,62 |
| Tibialis anterior | HFD | V_TCA_total | 4 | 1,69 | 0,261 | 1,648 | -0,139 | -7,62 |
| Tibialis anterior | Control | V_dil_tca | 2 | 0,12 | 0,17 | 0,12 | 0,015 | 12,548 |
| Tibialis anterior | HFD | V_dil_tca | 4 | 0,135 | 0,226 | 0,034 | 0,015 | 12,548 |
| Tibialis anterior | Control | V_ox_over_MR_glc | 2 | 0,594 | 0,124 | 0,594 | -0,171 | -28,799 |
| Tibialis anterior | HFD | V_ox_over_MR_glc | 4 | 0,423 | 0,097 | 0,444 | -0,171 | -28,799 |
| Tibialis anterior | Control | V_TCA_total_over_MR_glc | 2 | 0,636 | 0,065 | 0,636 | -0,15 | -23,565 |
| Tibialis anterior | HFD | V_TCA_total_over_MR_glc | 4 | 0,486 | 0,187 | 0,444 | -0,15 | -23,565 |